\documentclass[traditabstract]{aa} % for the abstract without structuration 
\usepackage{graphicx}
\usepackage{siunitx}
\usepackage{txfonts}
\usepackage{graphicx}
\usepackage{caption}
\usepackage{natbib}
\usepackage{subfigure} %um Bilder nebeneinander zu plotten
\usepackage{hyperref}
\usepackage[utf8]{inputenc}
\usepackage{placeins}
\usepackage{scrextend}
\usepackage{rotating} %Um die Tabellen 90 Grad zu drehen verwende
\usepackage{multirow} %for \multirow command
\usepackage{booktabs} % to use \toprule, \midrule and \bootmrule in table

\newcommand{\su}{sub-Ch}

\newcommand{\notshow}[1]{}

\newcommand{\bd}{\begin{displaymath}}
\newcommand{\ed}{\end{displaymath}}
\newcommand{\be}{\begin{equation}}
\newcommand{\ee}{\end{equation}}
\newcommand{\beaa}{\begin{eqnarray*}}
\newcommand{\eeaa}{\end{eqnarray*}}
\newcommand{\bea}{\begin{eqnarray}}
\newcommand{\eea}{\end{eqnarray}}

\newcommand{\erg}{\mathrm{erg}}
\newcommand{\s}{\mathrm{s}}

\newcommand{\cm}{\mathrm{cm}}

\newcommand{\lens}{D_\mathrm{d}}
\newcommand{\source}{D_\mathrm{s}}
\newcommand{\ls}{D_\mathrm{ds}}
\newcommand{\sourcez}{z_{\rm s}}
\newcommand{\lensz}{z_{\rm d}}
\newcommand{\lum}{{D_\mathrm{lum}}}

\newcommand{\Rein}{R_\mathrm{Ein}}

\newcommand{\dd}{\mathrm{d}}

\DeclareSIUnit\parsec{pc}
\DeclareSIUnit\lightyear{ly}
\DeclareSIUnit\year{yr}
\DeclareSIUnit\erg{erg}
\DeclareSIUnit\ster{ster}
\DeclareSIUnit\arcsec{arcsec}
\DeclareSIUnit\rad{rad}
\DeclareSIUnit\mag{mag}

\usepackage{color} %Text farbig machen
\definecolor{gruen}{cmyk}{0.35,0.01,0.80,0.1}

\definecolor{blue}{rgb}{0.0,0.0,1.0}
\definecolor{magenta}{rgb}{1.0,0.0,1.0}
\definecolor{orange}{rgb}{1.0,0.5,0.0}

\begin{document}
   \title{HOLISMOKES - III. Achromatic phase of strongly lensed type Ia supernovae}

  \titlerunning{Achromatic Phase of Strongly Lensed Type Ia Supernovae}

   \author{S.~Huber\inst{1,2}
   		 \and    
          S.~H.~Suyu\inst{1,2,3}
                    \and
          U.~M.~Noebauer \inst{1,4}  
                  \and 
             J.~H.~H.~Chan\inst{5}
             \and 
             M.~Kromer\inst{6}
             \and
             S.~A.~Sim\inst{7}
             \and
             D.~Sluse\inst{8}
             \and
             S.~Taubenberger\inst{1}
%SHS-waitMOconfirm%             M. Oguri\inst{9,10,11}
          }

   \institute{Max-Planck-Institut f\"ur Astrophysik, Karl-Schwarzschild Str. 1, 85741 Garching, Germany\\
              \email{shuber@MPA-Garching.MPG.DE}
         \and 
           Physik-Department, Technische Universit\"at M\"unchen, James-Franck-Stra\ss{}e~1, 85748 Garching, Germany
           \and
           Institute of Astronomy and Astrophysics, Academia Sinica, 11F of ASMAB, No.1, Section 4, Roosevelt Road, Taipei 10617, Taiwan
           \and
           Munich Re, IT1.6.4.1, Königinstraße 107, 80802 München
			\and
           Institute of Physics, Laboratory of Astrophysics, Ecole Polytechnique F\'ed\'erale de Lausanne (EPFL), Observatoire de Sauverny, 1290 Versoix, Switzerland
			\and
			Heidelberger Institut f\"ur Theoretische Studien,
Schloss-Wolfsbrunnenweg 35, 69118 Heidelberg, Germany
			\and
			Astrophysics Research Centre, School of Mathematics and Physics, Queen’s University Belfast, Belfast BT7 1NN, UK
			\and
			STAR Institute, Quartier Agora - Allée du six Août, 19c B-4000 Liège, Belgium
%SHS-waitMOconfirm%			\and
%SHS-waitMOconfirm%           Research Center for the Early Universe, University of Tokyo, 7-3-1 Hongo, Bunkyo-ku, Tokyo 113-0033, Japan
%SHS-waitMOconfirm%           \and
%SHS-waitMOconfirm%           Department of Physics, University of Tokyo, Tokyo 113-0033, Japan   
%SHS-waitMOconfirm%           \and
%SHS-waitMOconfirm%           Kavli Institute for the Physics and Mathematics of the Universe (Kavli IPMU, WPI), The University of Tokyo, Chiba 277-8582, Japan
}

   \date{Received --; accepted --}

% \abstract{}{}{}{}{} 
% 5 {} token are mandatory
 
  \abstract
  {%In the next years many strongly lensed Type Ia supernovae (LSNe Ia) will be detected. To use them for cosmology a time delay measurement between different images is necessary. Due to the sharp rise and decline SNe Ia light curves are promising for measuring time delays but microlensing can distort these light curves and therefore add large uncertainties to the measurements. An alternative approach are color curves where uncertainties due to microlensing are significantly reduced for a certain period of time. This is the so called achromatic phase and will be in detail investigated in this work. For this we have tested four different SNe Ia models on various different microlensing configurations. We find on average an achromatic phase of at least three rest-frame weeks in all color curves but the spread is quite large and an achromatic phase of just a few days is also possible. Further, the achromatic phase is longer for smoother microlensing maps, lower macro-magnifications and larger Einstein Radii where the latter sets the scale of the microlensing map. From our investigation we do not find a strong dependency on the model or on asymmetries in the SNe ejecta. Furthermore, we find for six rest-frame color curves features like extreme points within the achromatic phase which makes them promising for time delay measurements. These curves contain combinations of rest-frame bands \textit{u, g, r} and \textit{i} and to observe them for typical LSNe Ia redshifts it would be necessary to cover filters \textit{r, i, z, y, J,} and H. If follow-up resources are restricted we recommend \textit{r, i} and \textit{z} as bare minimum since LSNe Ia are bright there and observational uncertainties are lower as in the infrared regime, but with just 3 filters the usefulness of color curves for time delay measurements is questionable. 
To use strongly lensed Type Ia supernovae (LSNe Ia) for cosmology, a
time-delay measurement between the multiple supernova (SN) images is
necessary.  The sharp rise and decline of SN Ia light curves make them
promising for measuring time delays, but microlensing can distort
these light curves and therefore add large uncertainties to the
measurements. An alternative approach is to use color curves where
uncertainties due to microlensing are significantly reduced for a
certain period of time known as the achromatic phase. In this work, we
investigate in detail the achromatic phase, testing four different SN
Ia models with various microlensing configurations. We find on average
an achromatic phase of around three rest-frame weeks or longer for
most color curves, but the spread in the duration of the achromatic
phase (due to different microlensing maps and filter combinations) is
quite large and an achromatic phase of just a few days is also
possible. Furthermore, the achromatic phase is longer for smoother
microlensing maps and lower macro-magnifications.
From our investigations, we do not find a strong
dependency on the SN model or on asymmetries in the SN ejecta. We find
that six rest-frame LSST color curves exhibit features such as extreme
points or turning points within the achromatic phase, which make them
promising for time-delay measurements; however, only three of the
color curves are independent. These curves contain combinations of
rest-frame bands \textit{u, g, r,} and \textit{i,} and to observe them
for typical LSN Ia redshifts, it would be ideal to cover
(observer-frame) filters \textit{r, i, z, y, J,} and \textit{H}. If
follow-up resources are restricted, we recommend \textit{r, i,} and
\textit{z} as the bare minimum for using color curves and/or light
curves since LSNe Ia are bright in these filters and observational
uncertainties are lower than in the infrared regime. With additional
resources, infrared observations in \textit{y, J,} and \textit{H}
would be useful for obtaining color curves of SNe, especially at
redshifts above $\sim$ $0.8$ when they become critical.  }
   {}
   {}
   {}
   {}


   \maketitle
%
%________________________________________________________________

\section{Introduction}

On the one hand, there is a tension in the Hubble constant $H_0$ of at
least $4 \sigma$ \citep{Verde:2019ivm} between the early Universe
measurements \citep{Planck:2018vks} and late Universe measurements
from the Cepheids distance ladder 
\citep[e.g.,][]{Riess:2016jrr,Riess:2018byc,Riess:2019cxk}.  On the
other hand, the Hubble constant from the distance ladder using the tip
of the red giant branch \citep{Freedman:2019jwv,Freedman_2020} or surface brightness fluctuations for the calibration of Type Ia supernovae \citep[SNe Ia,][]{Khetan:2020hmh} is
consistent with both the Planck results and the Cepheids.  To assess
this tension and whether new physics is required to reconcile it, independent methods and measurements of $H_0$ are
important. Lensing time-delay cosmography is a powerful tool for
measuring $H_0$ in a single step \citep{Refsdal:1964}, independent of
other probes.  The time delay can be inferred from a variable source,
strongly lensed into multiple images by an intervening galaxy or
galaxy cluster. This technique has been applied successfully to six
lensed quasars to measure $H_0$ with 2.4 percent precision
\citep[e.g.,][]{Suyu:2016qxx,2017Courbin,Bonvin:2018dcc,Birrer:2018vtm,2019MNRAS.490..613S,Rusu:2019xrq,Chen:2019ejq,Wong:2019kwg},
and more systems are being analyzed
\citep{Shajib:2019toy,Millon:2019slk,Birrer+2020}.

Instead of quasars, strongly lensed type Ia supernovae (LSNe Ia) are
promising for measuring $H_0$ given that: (1) characteristic supernova
(SN) light curve shapes make time-delay measurements possible on
shorter time scales, (2) SNe fade away, facilitating measurements of the
dynamics of the lens
\citep{Barnabe2011,2017:Yildirim,Shajib:2018,Yildirim:2019vlv} to
break model degeneracies, for example the mass-sheet degeneracy
\citep{Falco:1985,Schneider:2013wga}, and (3) SNe Ia are
standardizable candles that allow us to break model degeneracies,
independently of dynamics, for lens systems whose lensing
magnifications are well characterized  
\citep{2003MNRAS.338L..25O,Foxley-Marrable:2018dzu}. So far two LSNe
with resolved multiple images have been observed, namely SN ``Refsdal"
\citep{Kelly:2015xvu,Kelly:2015vjq} and iPTF16geu
\citep{Goobar:2016uuf}, but we expect to find 500 to 900 LSNe Ia
\citep{Quimby:2014,GoldsteinNugent:2017,Goldstein:2017bny,Wojtak:2019hsc}
with the upcoming Rubin Observatory Legacy Survey of Space and Time
(LSST).

To measure time delays between different images of a LSN Ia, one could
use spectra, light curves and/or color curves. Problems arise from
microlensing; this phenomenon is similar to strong lensing but instead of galaxies and galaxy clusters, the compact objects (for example, stars) located in the main lens also deflect the light. Due to the low masses of the microlenses, multiple images are typically not resolvable and for these cases microlensing is only observable as additional magnification. Originally predicted by \cite{Chang_Refsdal_1979}, this phenomenon was unambiguously detected for the first time by \cite{Irwin:1989} in the quasar QSO 2237+0305 as uncorrelated brightness variations between the four multiple images. More information on quasar microlensing is available in, for example, \cite{2010GReGr..42.2127S} and \cite{2019BAAS...51c.487M}, and a general overview of microlensing is available in, for example, \cite{Wambsganss:2006nj} and \cite{2016aagl.book.....M}. For our case, the additional magnification from stars in the lens
galaxy influences images independently from one another and
therefore adds uncertainties to the delay measurement
\citep{Yahalomi:2017ihe,Goldstein:2017bny,Foxley-Marrable:2018dzu,Huber:2019ljb,
  PierelRodney+2019}.

While
the influence of microlensing on spectra and light curves is strong in
certain configurations, color curves have the following advantage: If
microlensing affects light curves from different filters in a similar
way, it cancels out in the color curves. This was first
investigated by \cite{Goldstein:2017bny}, who show that microlensed
color curves of LSNe Ia are ``achromatic," in other words, their color curves
are independent of microlensing for up to three rest-frame weeks after
explosion, and therefore time-delay uncertainties are reduced from
approximately $4\%$ to $1\%$ if color curves are used instead of light
curves. \cite{Huber:2019ljb} investigated this further using the
spherically symmetric W7 model \citep{1984:Nomoto} (similar to
\cite{Goldstein:2017bny}) for a specific magnification map (with
lensing convergence $\kappa$ = 0.6, shear $\gamma$ = 0.6, and smooth
matter fraction $s$ = 0.6). \cite{Huber:2019ljb} also find the
presence of an achromatic phase, but only for color curves where the
specific intensity profiles are similar.  In this paper, we explore
the achromatic phase further, notably to test if it is only present
in the W7 model or if other SN Ia models, including
multidimensional and asymmetric ones, also show an achromatic
phase. In addition, we investigate the dependency of the duration of the
achromatic phase on different microlensing maps and color curves.

This paper is organized as follows. In Section \ref{sec: SNe Ia
  models}, we present the different SN Ia models investigated in this
work. The calculation of microlensed SN Ia light curves is shown in
Section \ref{sec: Microlensing on Type Ia Supernova}, and results are
presented in Section \ref{sec: Achromatic phase of color curves}.
We conclude in Section \ref{sec: Discussion and summary}.

\section{SN Ia models and color curves}
\label{sec: SNe Ia models}
In Section \ref{sec: Theoretical SN Ia models} we give a short introduction to SNe Ia and the four theoretical models we use in this work. In Section \ref{sec: color curves of different SNe Ia models} we compare the color curves from the theoretical models to an empirical model.

\subsection{Theoretical SN Ia models}
\label{sec: Theoretical SN Ia models}
SNe Ia most likely have their origin in a thermonuclear explosion of a
carbon-oxygen white dwarf (WD) \citep[e.g.,][]{Hoyle:1960}, but
details about the progenitor and explosion mechanism are still
unknown. The classical textbook scenario is the single degenerate case
where a non-rotating WD is stable until it approaches the
Chandrasekhar mass limit, $ M_\mathrm{Ch} \approx 1.4 M_\odot$
\citep{Whelan:1973,Nomoto:1982}, due to the accretion from a main
sequence star or a red giant. Today this classical scenario has lost
some of its relevance and there are other mechanisms considered where
the WD explodes before the Chandrasekhar mass is reached, which are
typically called sub-Chandrasekhar (\su) explosions
\citep{Sim:2010}. Furthermore, for the thermonuclear burning, one
distinguishes between detonation (supersonic shock) and deflagration
(subsonic heat conduction). Another approach for SNe Ia is the violent
merger mechanism that belongs to the family of double-degenerate
scenarios in which the companion is another WD
\citep[see][]{Pakmor:2010,Pakmor:2011,Pakmor:2012,Pakmor:2013}. At the
moment it is not clear which of the scenarios is the right one to
describe SNe Ia and whether multiple scenarios or just a single one
can explain the observed SN Ia explosions. More details can be found
in, for example, \cite{Hillebrandt:2000}, \cite{Hillebrandt:2013gna}, and \cite{Livio:2018rue}.

In this work we investigate four different SN Ia models to test the
dependency of the achromatic phase on these models. The first one is
the W7 model \citep{1984:Nomoto}, a (parameterized) 1D deflagration of
a carbon-oxygen white dwarf close to the Chandrasekhar mass
($M_\mathrm{Ch}$ CO WD) with $0.59 M_\odot$ of $^{56}$Ni, which is
known to reproduce key observables of normal SNe Ia, for example, spectra \citep{Jeffery:1992,Nugent_1997,Baron_2006,Gall_2012}.  Furthermore, we investigate
the N100 model \citep{Seitenzahl:2013}, which is a delayed detonation
model, in other words, the burning starts as 3D deflagration and transitions
into a detonation later, of a $M_\mathrm{Ch}$ CO WD. In this
particular delayed detonation model, the explosion produced $0.6
M_\odot$ of $^{56}$Ni.  The third model is a \su\, model with a
carbon-oxygen WD of $1.06 M_\odot$ producing $0.56 M_\odot$ of
$^{56}$Ni \citep{Sim:2010}.  In addition, we consider a merger model
from \cite{Pakmor:2012} where two carbon-oxygen WDs of $0.9 M_\odot$
and $1.1 M_\odot$ collide and ignite a detonation in which $0.62
M_\odot$ of $^{56}$Ni are produced.

\begin{figure}
\centering
\includegraphics[width = 0.42 \textwidth]{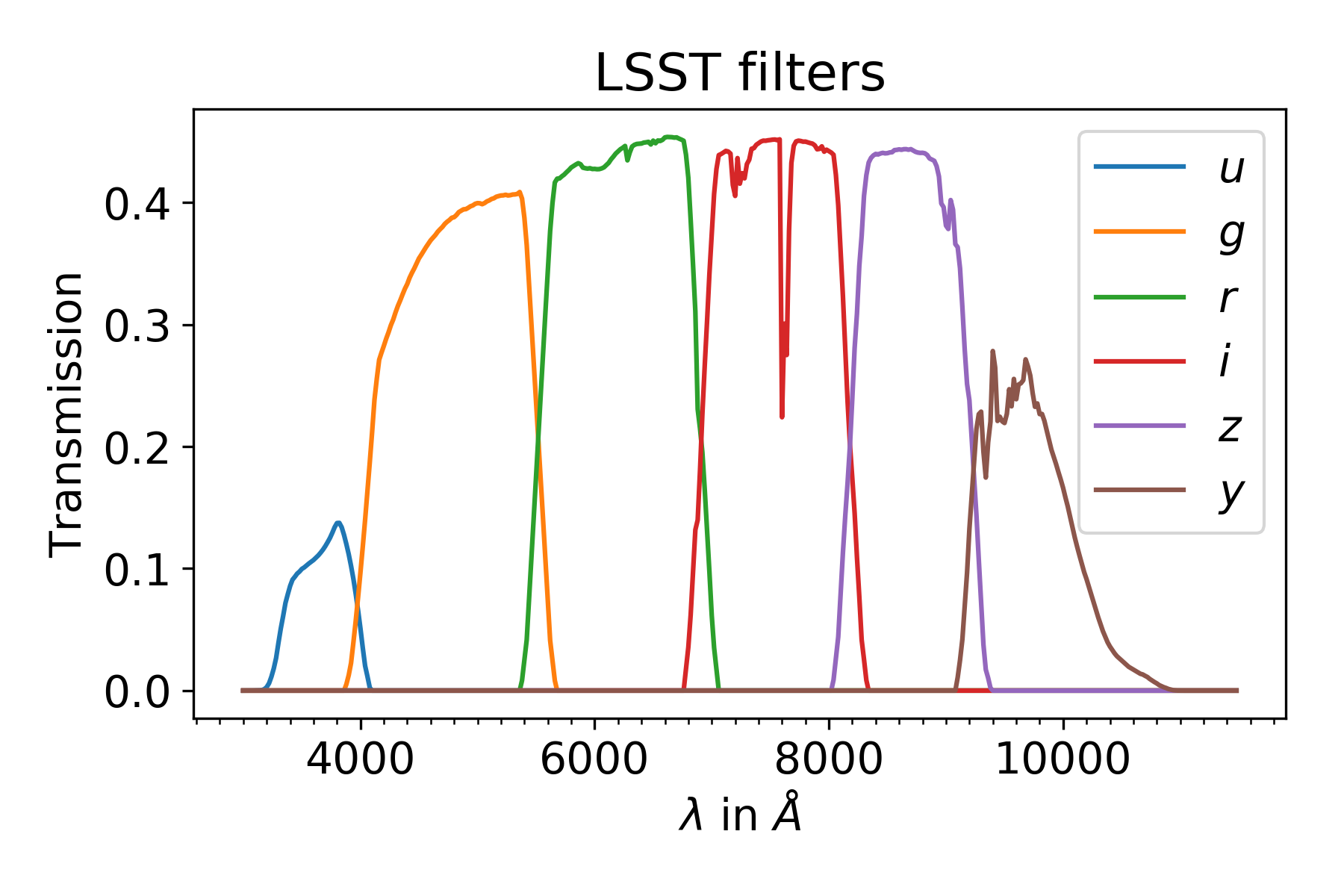}

\caption{Transmission for all six LSST filter bands. The effective
  wavelengths are $\lambda_\mathrm{eff,\textit{u}} =
  \SI{3671}{\angstrom}, \lambda_\mathrm{eff,\textit{g}} =
  \SI{4827}{\angstrom}, \lambda_\mathrm{eff,\textit{r}} =
  \SI{6223}{\angstrom}, \lambda_\mathrm{eff,\textit{i}} =
  \SI{7546}{\angstrom}, \lambda_\mathrm{eff,\textit{z}} =
  \SI{8691}{\angstrom},$ and $\lambda_\mathrm{eff,\textit{y}} =
  \SI{9712}{\angstrom}$.}
\label{fig:LSST filters transmission functions}
\end{figure}

From our theoretical SN models we can get the observed flux $F_{\lambda,\mathrm{o}}(t)$ from which we can calculate AB magnitudes following
\begin{equation}
\scalebox{1.13}{$
m_\mathrm{AB,X}(t) = -2.5 \log_{10} \left(\frac{\int \dd\lambda \, \lambda S_\mathrm{X}(\lambda) \, F_{\lambda,\mathrm{o}}(t)  }{\int \dd \lambda \, S_\mathrm{X}(\lambda) \, c \, / \lambda} \times \si{\square\cm\over\erg} \right)  - 48.6$}
\label{Eq: microlensed light curves for ab magnitudes}
\end{equation}
\citep{Bessel:2012}, where $c$ is the speed of light and
$S_\mathrm{X}(\lambda)$ is the transmission function\footnote{\url{https://github.com/lsst/throughputs/tree/master/baseline}} for the LSST
filter X (that can be \textit{u}, \textit{g}, \textit{r}, \textit{i},
\textit{z}, or \textit{y}), which are illustrated in Figure
\ref{fig:LSST filters transmission functions}.

\subsection{Color curves of different SN Ia models}
\label{sec: color curves of different SNe Ia models}

\begin{figure*}
\centering
\includegraphics[width = 0.98 \textwidth]{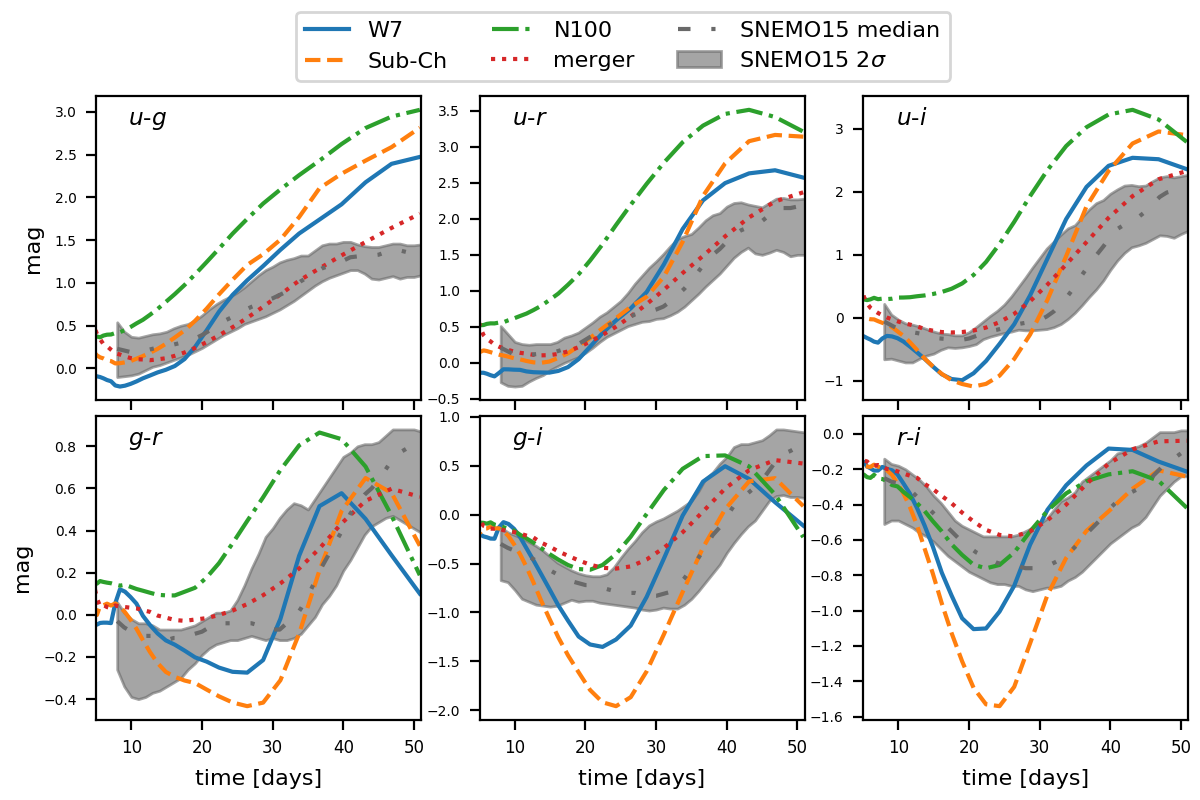}
\caption{Comparison of the four theoretical models: W7, N100, \su, and the merger to the empirical model \texttt{SNEMO15}.}
\label{fig: SNEMO 15 color curves comparsion to models}\end{figure*}

In this work we are especially interested in color curves and the dependence of the achromatic phase on the assumed theoretical SN Ia model. Therefore, we first compare in 
Figure \ref{fig: SNEMO 15 color curves comparsion to models} the color curves of the SN Ia models used in this work to the empirical SN Ia model \texttt{SNEMO15} from \cite{Saunders:2018rjn} for six LSST color curves. We pick \texttt{SNEMO15} instead of \texttt{SNEMO2} or \texttt{SNEMO7} as this provides the largest variety in colors and therefore represents best the scatter of the colors based on real observations.
To produce the median and $2\sigma$ (97.5th percentile $-$ 2.5th percentile) curves of \texttt{SNEMO15}, we consider all 171 SNe Ia from \cite{Saunders:2018rjn} used for training and validation of the empirical SN model.
Furthermore, we use the flux from the empirical model to calculate the LSST color curves using Equation (\ref{Eq: microlensed light curves for ab magnitudes}).
We note that the data cover only $ \SI{3305}{\angstrom}$ to $\SI{8586}{\angstrom}$, and therefore color curves containing the \textit{z} and \textit{y} bands cannot be calculated.
Considering the \textit{u} band, we find that the filter transmission becomes relevant around $\SI{3200}{\angstrom}$. 
Therefore, our presented \textit{u} band light curve is an approximation, but a reasonably accurate one since flux drops further in the UV region at short wavelengths and
the LSST filter transmission in the missing $\SI{100}{\angstrom}$ region is low. 

The comparison of the W7 model from this work to the one
presented in \cite{Goldstein:2017bny}, where a different radiative
transfer code has been used, shows a different trend in the color
evolution. 
Details about uncertainties in the light curve shapes due
to different ionization treatments in \texttt{ARTIS} can be found in \cite{Kromer:2009ce}. 
A spectral comparison of different SN Ia models shows that differences are larger 
in early stages ($\sim$8 days) in comparison to later epochs ($\sim$30 days),
but \cite{Suyu:2020opl} point out as well that the exact spectral shapes depend on approximations that are used in the radiative transfer calculations \citep{Dessart+2014,Noebauer:2017vsf}.

As we see in Figure \ref{fig: SNEMO 15 color curves comparsion to models}, none of the theoretical models are able to predict the color curves perfectly, but theses models are required to investigate impacts of microlensing on color curves and it is particularly interesting, how the microlensing signal differs between models.  
Furthermore, even though the values of the color curves for
different models are offset, the overall color evolution do match the trend of \texttt{SNEMO15}. 
Although features are stronger in the theoretical models, they are still present in the color curves from \texttt{SNEMO15}. To measure time delays from color curves,
features like extreme points or turning points would be crucial, to mitigate color differences between images due to differential dust extinction \citep{Eliasdottir:2006gd}. 
Furthermore, these extreme or turning points should be located within the achromatic
phase, to reduce uncertainties due to microlensing. 

%Even though the values of the color curves for
%different models are offset, the overall color evolution do match and
%show a similar trend as the observed SN2011fe \citep{Nugent:2011},
%which allows to draw conclusions from our models. LSST color curves
%for SN2011fe have been calculated from spectra of the
%SNfactory\footnote{\url{https://snfactory.lbl.gov/index.html}}
%\citep{Aldering:2002, Pereira:2013fas} using Equation (\ref{Eq:
%  microlensed light curves for ab magnitudes}). We note that data from
%the SNfactory covers only $ \SI{3300}{\angstrom}$ to $\SI{9700}{\angstrom}$ and therefore color curves containing the
%\textit{y} band cannot be calculated. Considering the \textit{u}
%band, we find that the filter transmission becomes relevant around
%$\SI{3200}{\angstrom}$. Therefore our presented \textit{u} band light
%curve is an approximation, but a reasonably accurate one since flux
%drops further in the UV region at short wavelengths and the LSST
%filter transmission in the missing $\SI{100}{\angstrom}$ region is low.

\section{Microlensing on SNe Ia}
\label{sec: Microlensing on Type Ia Supernova}
The calculation of microlensing on SNe Ia, which we use in this work, is described in detail by
\cite{Huber:2019ljb}, who where motivated by the work of
\cite{Goldstein:2017bny}. While both assumed the W7 model for the SNe Ia, \cite{Huber:2019ljb} calculated synthetic observables using the
radiative transfer code \texttt{ARTIS} \citep{Kromer:2009ce} whereas
\cite{Goldstein:2017bny} used \texttt{SEDONA} \citep{Kasen:2006ce}. In
the following, we briefly summarize the general idea.

To calculate microlensed light curves, we use the emitted specific
intensity $I_{\lambda,e}(t,p)$ at the source plane calculated via
\texttt{ARTIS} for a given SN model, where $I_{\lambda,e}(t,p)$ is a
function of wavelength $\lambda$, time since explosion $t$, and impact
parameter $p$, which is the projected distance from the ejecta center. We
combine $I_{\lambda,e}(t,p)$ with magnification maps from {\tt
  GERLUMPH} \citep[J.~H.~H.~Chan in preparation]{Vernardos:2015wta},
which uses the inverse ray-shooting technique
\citep[e.g.,][]{Kayser:1986,Wambsganss:1992,Vernardos:2013vma}
yielding the magnification factor $\mu(x,y)$ as a function of
cartesian coordinates $x$ and $y$ on the source plane\footnote{We note
  that $\mu$ denotes the magnification factor and not $\cos \theta$ as
  usually in radiative transfer equations.}.

Throughout this work, the specific intensity is treated in spherical
symmetry and has therefore just a 1D spatial dependency on $p$. This
approximation is exact for the W7 model and the \su\, model of
\cite{Sim:2010}, and good for the N100 model, which produces nearly
spherically symmetric ejecta, but results inferred from the asymmetric
merger model are questionable. To estimate the impact of viewing angle
effects we investigate the asymmetries in the merger model by using
only photons from one half of the ejecta, for example, only averaging over
photons that emerge in positive $x$-direction.

Magnification maps are determined by three main parameters, namely the
convergence $\kappa$, the shear $\gamma$, and the smooth matter
fraction $s=1-\kappa_*/\kappa$, where $\kappa_*$ is the convergence of
the stellar component. Furthermore, we assume a Salpeter initial mass
function with a mean mass of the point mass microlenses of $\langle M
\rangle = 0.35 M_\odot$. Details of the initial mass function have
negligible impact on our studies (J.~H.~H.~Chan in preparation). 
Our maps have
a resolution of 20000 $\times$ 20000 pixels with a total
size\footnote{As a cross-check we investigated for a few cases
  larger maps with a total size of $20 \Rein$ $\times$ $20 \Rein$, and
  our conclusions drawn in this work do not depend on the size of the
  maps.} of $10 \Rein$ $\times$ $10 \Rein$. The Einstein Radius is a
characteristic size of the map and can be calculated via
\begin{equation}
\Rein=\sqrt{\frac{4 G \langle M \rangle}{c^2} \frac{\source \ls}{\lens}},
\label{basics: Einstein Radius physical coordinate in cm}
\end{equation}
where $\source$, $\lens$ and $\ls$ are the angular diameter distances
from the observer to the source, from the observer to the lens
(deflector), and between the lens and the source, respectively. To
calculate these distances we assume a flat $\Lambda$CDM cosmology
where we neglect the contribution of radiation ($H_0 =
\SI{72}{\km\per\s\per\mega\parsec}$ and $\Omega_m = 0.26$ as assumed
by \citet[][hereafter OM10]{Oguri:2010}, our reference for
typical LSN Ia images used in this work).

To calculate the observed microlensed flux we place the SNe Ia in the
magnification map and solve:
\begin{equation}
F_{\lambda,\mathrm{o}}(t)=\frac{1}{\lum^2(1+\sourcez)}\int \dd x \int \dd y \, I_{\lambda,\mathrm{e}}(t,x,y) \, \mu(x,y),
\end{equation}
where $\lum$ is the luminosity distance to the source. Furthermore, we
interpolate the emitted specific intensity $I_{\lambda,\mathrm{e}}$
onto a 2D cartesian grid ($x,y$) with the same spatial resolution as
the microlensing map and integrate over the whole size of the SN
\citep{Huber:2019ljb}, which depends on the time after explosion. We
then obtain the AB-magnitudes via Equation (\ref{Eq: microlensed light curves for ab magnitudes}).

\begin{figure}
\centering
\includegraphics[width = 0.42 \textwidth]{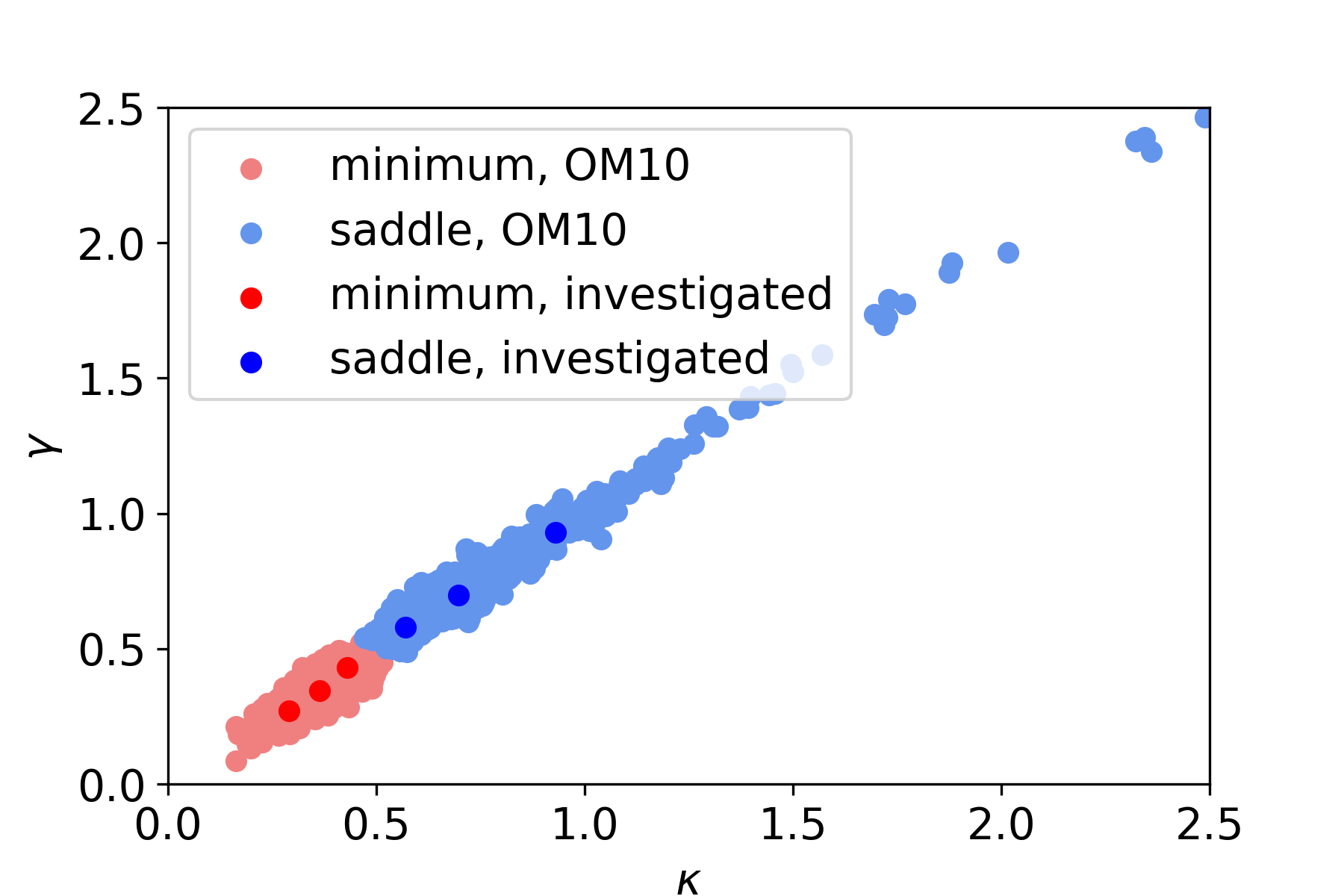}
\caption{Distribution of $\kappa$ and $\gamma$ values for two
  different image types (minimum and saddle, in red and blue,
  respectively), showing the sample of the OM10 catalog and the six
  pairs we have investigated ($\kappa, \gamma = (0.29, 0.27), (0.36,
  0.35), (0.43, 0.43), (0.57, 0.58), (0.70, 0.70),$ and $(0.93,
  0.93)$). There are a few saddle points not shown in the plot going
  up to $\kappa$ and $\gamma$ values of around 10.}
\label{fig: kappa gamma distribution}
\end{figure}

\section{Achromatic phase of LSNe Ia}
\label{sec: Achromatic phase of color curves}
In this section, we investigate the dependency of the achromatic phase
on different parameters.  Section \ref{sec: SNe Ia models achromatic
  phase investigation} probes the four different SN Ia models (W7, N100,
 \su \, model, and the merger models) under the assumption of
spherical symmetry. In addition, the dependency on the smooth matter
fraction $s$ and different image configurations (varying $\kappa$ and
$\gamma$) is discussed. Section \ref{sec: Redshift Dependency} shows
the dependency on the scale of the microlensing map defined via
$\Rein$, which depends on the source and lens redshifts, $\sourcez$
and $\lensz$, respectively. In Section \ref{sec: Asymetric Merger
  Model}, effects of asymmetries in the merger model are
investigated. While Section \ref{sec: SNe Ia models achromatic phase
  investigation} to \ref{sec: Asymetric Merger Model} assume
rest-frame color curves, redshifted color curves are investigated in
Section \ref{sec:Redshifted color curves}. In Table
\ref{tab:SummaryPars}, we summarize the number of models and different
magnification maps that are taken into account in the different
sections.

\begin{table}
\tabcolsep=0.13cm
\begin{tabular}{ccccccc}
\hline
Section & models & asymmetries & $\sourcez, \lensz$ & $\kappa, \gamma$ & $s$ & color curves \\
\hline
\ref{sec: SNe Ia models achromatic phase investigation} & 4 & no &1 & 6 & 5 & 15 \\
\ref{sec: Redshift Dependency} & 4 & no & 4 & 2 & 5 & 15 \\
\ref{sec: Asymetric Merger Model} & 1 & 6 & 1 & 2 & 5 & 15 \\
\ref{sec:Redshifted color curves} & 4 & no & 3 & 2 & 5 & 15 \\
\hline
\end{tabular}
\caption{\label{tab:SummaryPars}Summary of the number of different parameters investigated in Sections \ref{sec: SNe Ia models achromatic phase investigation} to \ref{sec:Redshifted color curves}. In Section \ref{sec: SNe Ia models achromatic phase investigation} the dependency on different models, smooth matter fractions and image configurations is investigated. Section \ref{sec: Redshift Dependency} exhibits the dependency on the scale of the magnification map, Section \ref{sec: Asymetric Merger Model} contains the investigations of asymmetries for the merger model, and Section \ref{sec:Redshifted color curves} shows redshifted color curves.}
\end{table}

\subsection{SN Ia models, smooth matter fraction, and image configuration}
\label{sec: SNe Ia models achromatic phase investigation}

To investigate the achromatic phase of LSNe Ia, we pick typical lens
and image configurations from the OM10 catalog \citep{Oguri:2010}. For
the source and lens redshifts we assume $\sourcez = 0.77$ and $\lensz
= 0.32$, which are the median values of the OM10 catalog. We use
redshifts in Sections \ref{sec: SNe Ia models achromatic phase
  investigation} to \ref{sec: Asymetric Merger Model} only to
calculate the scale of the microlensing maps, hence $\Rein$, and
therefore the color curves discussed in these sections are in the
rest-frame. The investigated $\kappa$ and $\gamma$ values are also
based on OM10 and shown in Figure \ref{fig: kappa gamma distribution}
as six dark points for two different image types (minimum and saddle, in
red and blue, respectively). For each of the two image types, the
investigated points correspond to the median values and the 16th and
84th percentiles of the OM10 sample, taken separately for $\kappa$ and
$\gamma$. For each of the six pairs of $\kappa$ and $\gamma$, five
different $s$ values (0.1, 0.3, 0.5, 0.7, and 0.9) are considered, which
cover typical $s$ values at image positions of galaxy-scale lenses
\citep[e.g.,][]{Schechter2014,Chen2018,Bonvin:2019xvn}.
%SHS1103:added
Therefore, we have in total 30 magnification maps (from $6\times5$).  Magnification
maps for most configurations we investigate in this work are shown in
Appendix \ref{sec: Microlensing maps}.

In comparison to the 30 different magnification maps probed in this work, 
\cite{Huber:2019ljb} have analyzed only a single magnification map ($\kappa, \gamma, s = 0.6$). \cite{Goldstein:2017bny} investigated a much larger sample of LSNe Ia with 78184 multiple images, but the sample is dominated by small-image-separation systems that will not be resolvable from ground-based monitoring, whereas we focus on spatially resolvable systems for cosmography by using the OM10 sample of strong lenses. Although our sample of LSNe Ia is much smaller, we probe per map 10000 random positions instead of just one as by \cite{Goldstein:2017bny}; therefore, we investigate a much larger sample of microlensed SN Ia color curves, which further allows us to probe dependencies such as the duration of the achromatic phase on different image configurations.

\begin{figure*}[htbp]
\centering
\subfigure{\includegraphics[trim=10 10 10 10,clip,width=.38\textwidth]{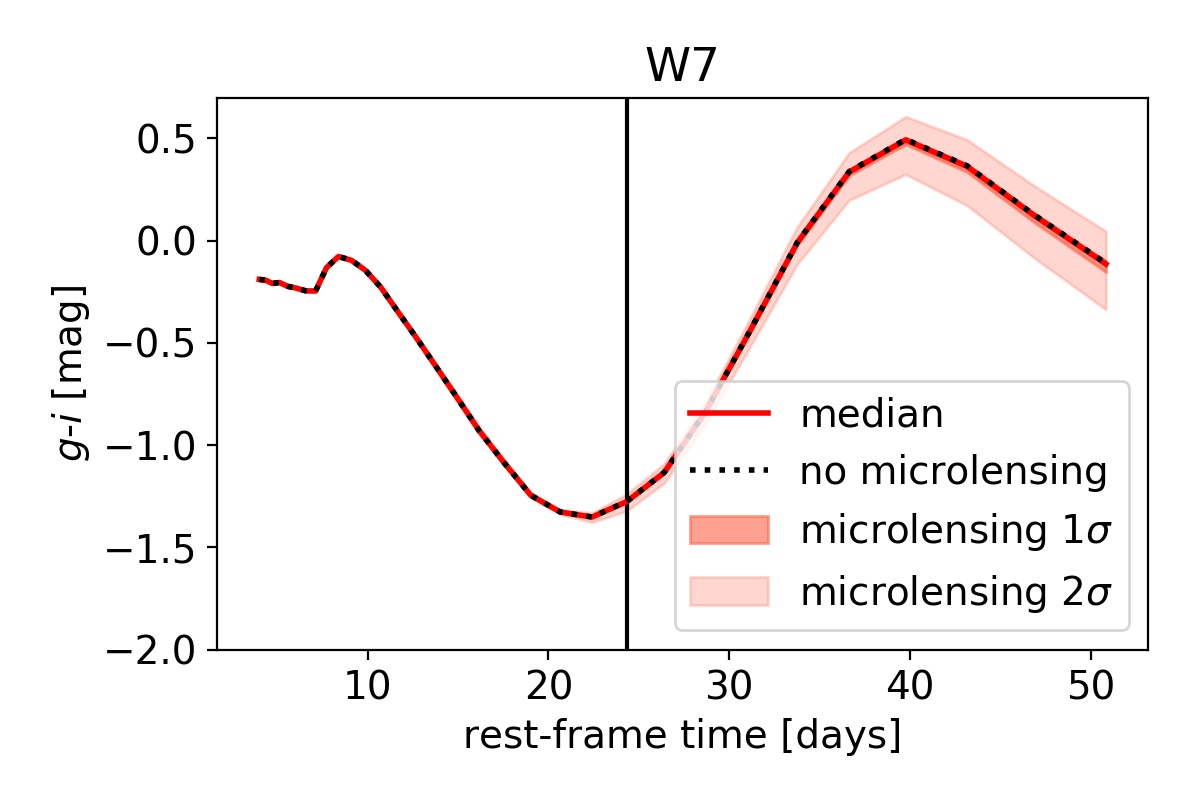}}
\subfigure{\includegraphics[trim=10 10 10 10,clip,width=.38\textwidth]{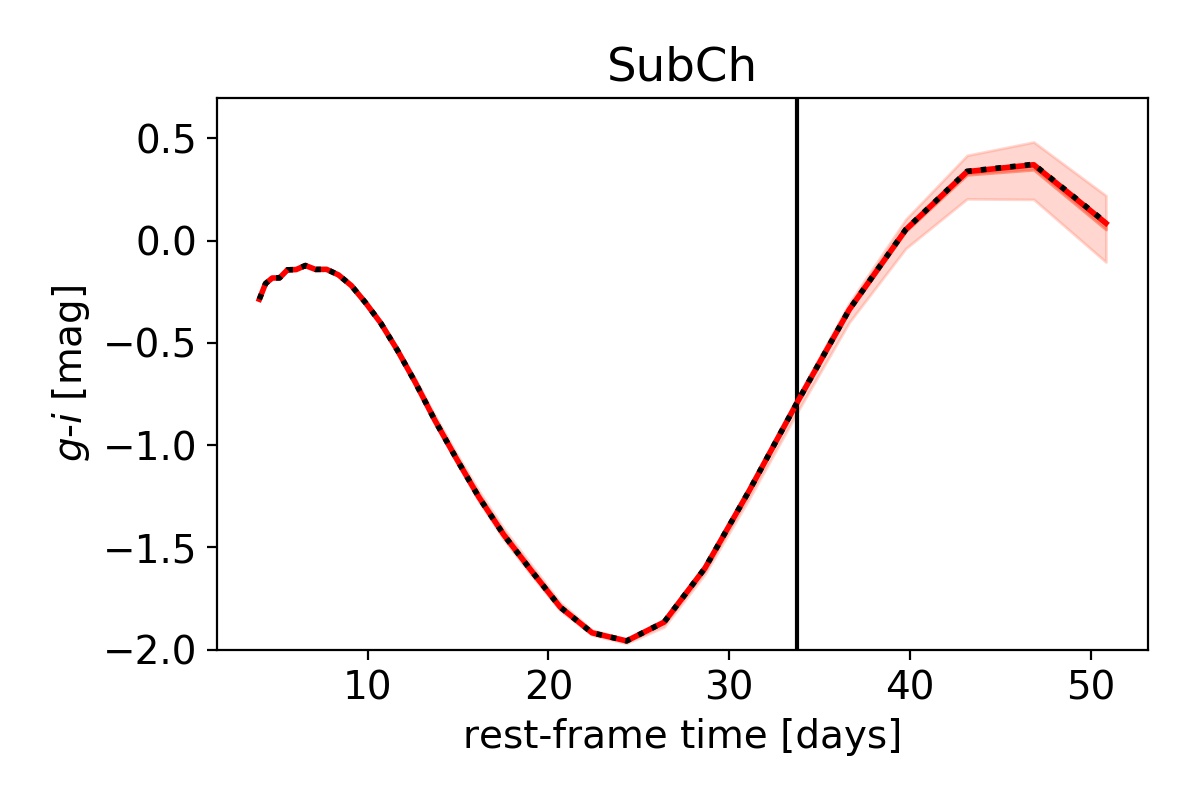}}
\subfigure{\includegraphics[trim=10 10 10 10,clip,width=.38\textwidth]{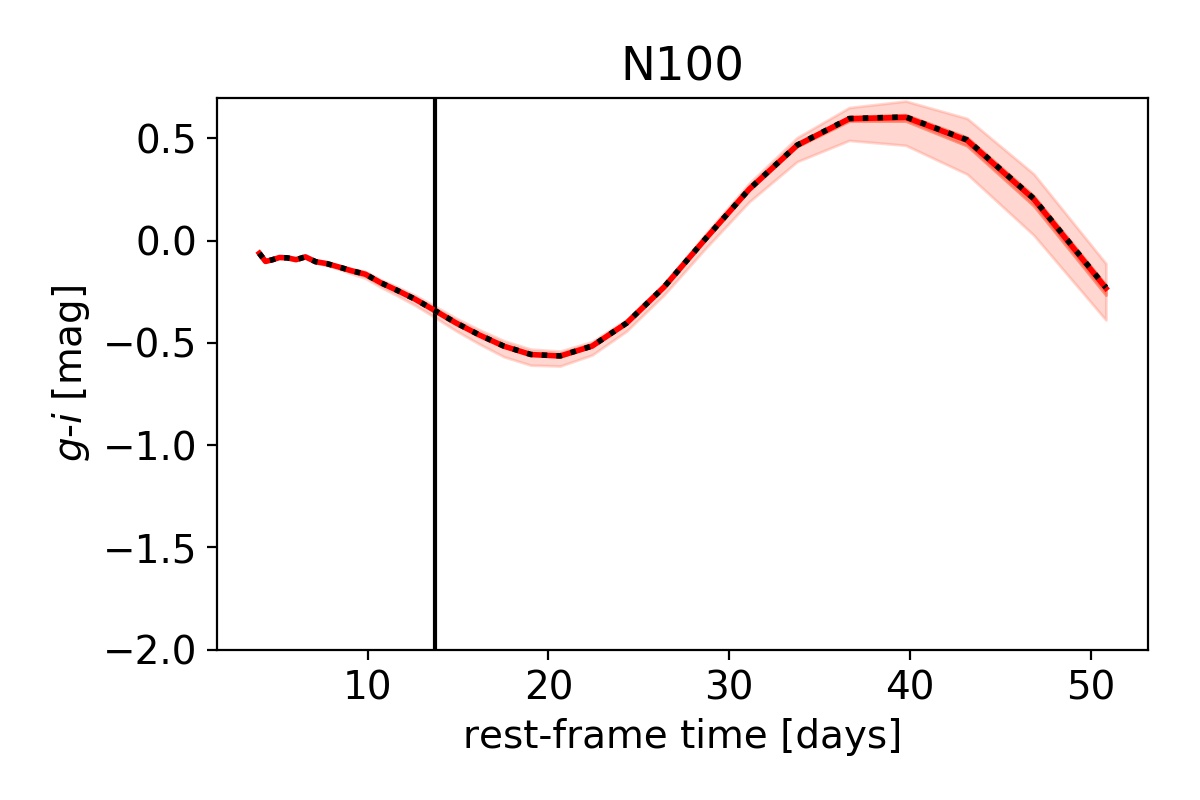}}
\subfigure{\includegraphics[trim=10 10 10 10,clip,width=.38\textwidth]{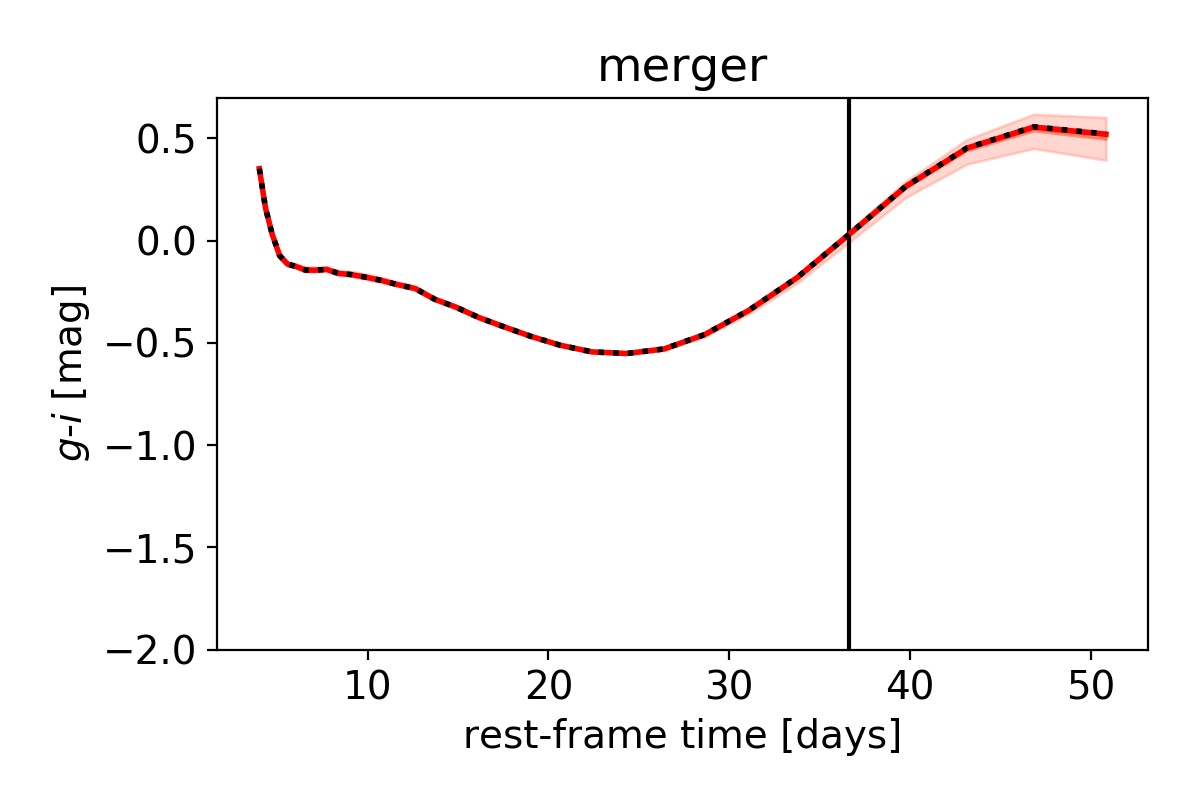}}
\caption{Rest-frame LSST color curves for 10000 random SN positions in
  the magnification map with $\kappa = 0.36, \gamma = 0.35$ and $s=
  0.5$, comparing microlensed color curves (with median in solid red,
  and 1$\sigma$ and 2$\sigma$ band in different shades) to
  non-microlensed ones (dotted black). The vertical black lines
  indicate the duration of the achromatic phase. We find different
  durations of the achromatic phase for different SN Ia models, where
  the N100 model in this specific case has the shortest and the merger
  model the longest duration. This is just a specific case to
  illustrate the dependency of the achromatic phase on the specific
  intensity profiles. More general conclusions can be drawn from
  Figures \ref{fig:color curves from LSST for different supernova
    model} and \ref{fig: achromatic phase as function of model, s and
    kappa gamma}.}
\label{fig: microlensed sigma color curves}
\end{figure*}

In our analysis, we draw for each of the 30 magnification maps 10000 random positions, where we calculate for each
position all six LSST microlensed light curves following Equation
(\ref{Eq: microlensed light curves for ab magnitudes}) and from these
the 15 LSST color curves (obtained through all possible pair-wise
combinations of the light curves). For each color curve, we consider
the $1\sigma$ band (``full-width," corresponding to the difference
between the 84th percentile and 16th percentile) and $2\sigma$ band
(full-width, corresponding to the difference between the 97.5th
percentile and 2.5th percentile) from the 10000 random positions,
which is shown in Figure \ref{fig: microlensed sigma color curves} for
\textit{g-i} for $\kappa = 0.36, \gamma = 0.35$ and $s= 0.5$ for all 4
SN models investigated in this work. We
use 32 time bins, covering rest-frame days 4.2 to
50.8 after explosion of the SNe Ia.

The black vertical lines in Figure \ref{fig: microlensed sigma color
  curves} correspond to the duration of the achromatic phase
$t_\mathrm{achro}$ which we define as the first of two neighboring
time bins where the 2$\sigma$ band becomes larger than the threshold
\begin{equation}
\Delta_{\mathrm{threshold}} = \mathrm{max}(0.05 \, \mathrm{mag},\Delta_{3 \%}),
\label{eg:Delta Threshold for duration of achromatic phase}
\end{equation}
where $\Delta_{3 \%} = 0.03 \, (\mathrm{max (color \, curve)}
- \mathrm{min (color \, curve)}$).  The reasoning behind these
definitions is as follows.  Typically the 2$\sigma$ band increases with time
but for rare cases it can exceed $\Delta_{\mathrm{threshold}}$ for a
single time bin and then drop below that limit again. With our
definition of $t_\mathrm{achro}$, including more than one time bin
that exceeds $\Delta_{\mathrm{threshold}}$, we skip such
outliers.  If multiple time bins are within one
day we require that all time bins exceed the
$\Delta_{\mathrm{threshold}}$ in the 2$\sigma$ band and set
$t_\mathrm{achro}$ as the lowest time bin\footnote{Only relevant for
  first few time bins since time is binned
  logarithmically.}.

To motivate the $0.05 \, \mathrm{mag}$ criterion, we look at a LSNe Ia
at the median source redshift of OM10, $\sourcez = 0.77$, and assume as
an image the median minimum corresponding to $(\kappa,\gamma) =
(0.36,0.35)$ leading to a macro-magnification of $3.5$. According to
\cite{Huber:2019ljb}, follow-up observations going between 1 and
$\SI{2}{\mag}$ deeper than LSST is ideal to measure time delays of
LSNe Ia. Assuming observations of around $\SI{1.5}{\mag}$ deeper than
the LSST-like 5$\sigma$ depth, we expect the 2$\sigma$ uncertainties
at the light curve peak of around $\SI{0.025}{\mag}$ in \textit{r} and
\textit{i} bands and higher uncertainties in other bands
\citep{2009:LSSTscience}. This 2$\sigma$ uncertainty corresponds to
$\SI{0.05}{\mag}$ in the $2 \sigma$ band, and therefore our first limit
in Equation (\ref{eg:Delta Threshold for duration of achromatic
  phase}) assures that microlensing uncertainties within the
achromatic phase are smaller than typical observational uncertainties. This is
a very conservative estimate, especially for color curves having much
larger changes than $\SI{0.05}{\mag}$ over time. This motivates the
use of $\Delta_{3 \%}$ as the second criterion in Equation
(\ref{eg:Delta Threshold for duration of achromatic phase}). This
means that for color curves covering more than ${\sim}\SI{1.7}{\mag}$
over time the $\SI{0.05}{\mag}$ criterion is replaced by the slightly
larger value $\Delta_{3 \%}$. Variations of color curves are always
within $ \SI{3}{\mag}$ so uncertainties due to microlensing are kept
within $\SI{0.09}{\mag}$ in the achromatic phase.

As an illustration, the \textit{g-i} color curve in the
upper-right-hand panel (\su) of Figure \ref{fig: microlensed sigma
  color curves} would have an achromatic phase of
${\sim}\SI{15}{\day}$ with only the first criterion in Equation
(\ref{eg:Delta Threshold for duration of achromatic phase}), but an
achromatic phase of ${\sim}\SI{35}{\day}$ with both requirements,
which is more appropriate given the large color variation. As a cross-check,
we compared different definitions of the achromatic phase,
namely, using only $\Delta_{\mathrm{threshold}} = 0.05$ or taking the
mean of the first two time bins instead of requiring multiple
neighboring bins. Although this changes the duration of the achromatic
phase for some cases, our general conclusions from averaging over many
different microlensing maps, color curves or models, are not
influenced. While our definition of the achromatic phase is arbitrary
to some extent, it is justified by looking at many plots as shown
in Figure \ref{fig: microlensed sigma color curves}, and we keep the
definition consistent over the whole work, which allows us to
compare different models and microlensing parameters.
%This is necessary because otherwise a color curve like \textit{g-i} shown in the upper right panel of Figure \ref{fig:color curves from LSST for different supernova model} would have a much shorter duration of the achromatic time (${\sim}\SI{15}{\day}$ instead of ${\sim}\SI{35}{\day}$), given that the assumed spread of $\SI{0.05}{\mag}$ is still very small in comparison to the total range of the color evolution. 

In Figure \ref{fig: microlensed sigma color curves}, we see that the
duration of the achromatic phase for this example depends on the SN Ia
model. While the model N100 has a short achromatic phase of
${\sim}\SI{15}{\day}$, the merger and \su\, models show a
significantly longer duration of ${\sim}\SI{35}{\day}$ and the W7
model is in between with ${\sim}\SI{25}{\day}$. This is related to the
specific intensity profiles, shown in Figure \ref{fig:specific
  intenstiy profiles for different supernova model}, for the radial
distribution of the radiation in the six LSST filters. The influence of
different specific intensity profiles on spectra and light curves is
presented in detail in Appendix A of \cite{Huber:2019ljb}.

The profiles for filters \textit{g} and \textit{i} for the \su \,
model and N100 at day 14.9 show larger differences than W7 and the
merger model, explaining the much shorter achromatic phase for the
N100 model. The \su \, model reaches the $\SI{0.05}{\mag}$ at a
similar time as the N100 model but given the large range covered by
the color evolution, the achromatic phase of the \su \, model is much
longer. In the case of the merger model, the long achromatic time is
explained by the intensity profiles because for later times (day
33.8), only the merger model still shows quite similar intensity
profiles in \textit{g} and \textit{i} explaining the longer duration
of the achromatic phase in the color curve \textit{g-i}.

\begin{figure*}
\centering
\includegraphics[width = \textwidth]{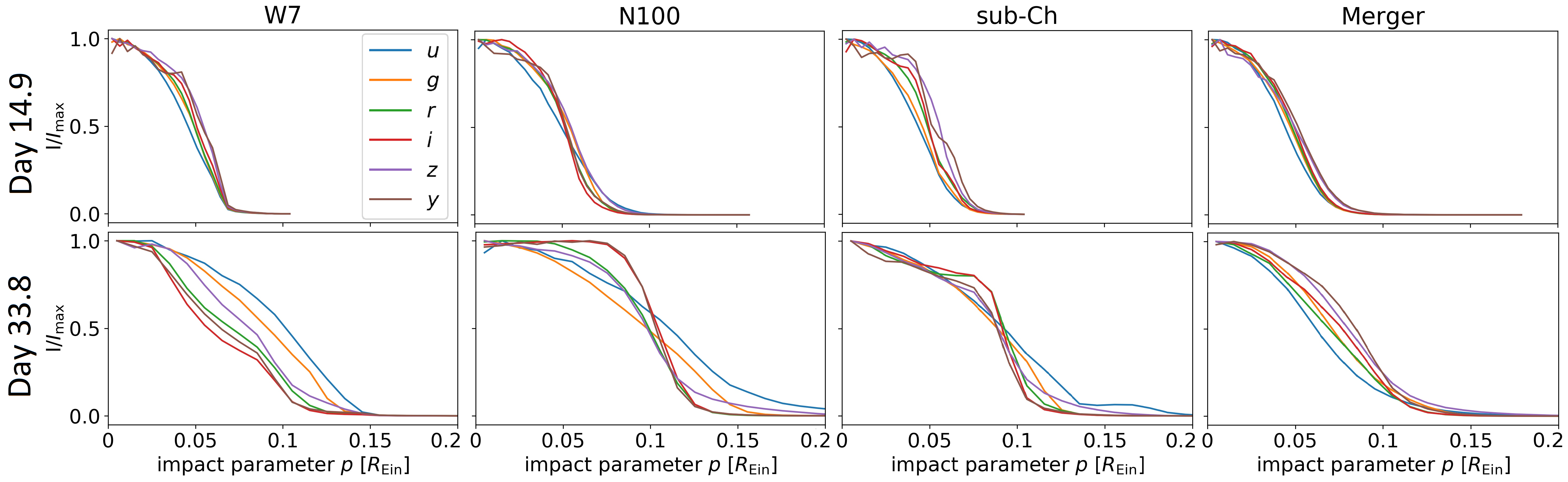}
\caption{Radial specific intensity profiles for four SN Ia models (W7,
  N100, the \su, and the merger model in the four labeled columns) for
  the six LSST filters at rest-frame day 14.9 (top row) and day 33.8
  (bottom row) after explosion, where $\Rein = 2.9 \times 10^{16} \,
  \mathrm{cm}$ for the median source and lens redshifts $(\sourcez,
  \lensz = 0.77,0.32)$ based on OM10. Different SN Ia models yield
  different specific intensity profiles, leading to different
  durations of the achromatic phase. The wiggles for low $p$ values at
  Day 14.9 are most likely due to Monte Carlos noise and we do not
  expect them to be physical. For our microlensing
  calculations, differences at higher $p$ values are more relevant,
  since events where micro caustics are crossed become more likely
  with larger radius.}
\label{fig:specific intenstiy profiles for different supernova model}
\end{figure*}

To draw a more general conclusion, we look at all color curves for the
6 $\times$ 5 magnification maps (see Table \ref{tab:SummaryPars}) for
different SN Ia models as shown in Figure \ref{fig:color curves from
  LSST for different supernova model} and \ref{fig:color curves from
  LSST for different supernova model not useful}. For each of the 30 magnification maps, we
have per color curve and model a $t_\mathrm{achro}$ from the 10000
random positions as shown in Appendix \ref{sec:Appendix achromatic
  phase in detail}. The vertical lines in Figure \ref{fig:color curves
  from LSST for different supernova model} and \ref{fig:color curves from
  LSST for different supernova model not useful} mark the mean values from
the 30 $t_\mathrm{achro}$, and are matched in color and linestyles to that
of the SN Ia models. In Figure \ref{fig:color curves from LSST for different supernova
  model}, six color curves are shown, which we refer to as useful color curves; these
color curves have, for at least three SN models, substantially non-linear
features like extreme points and turning points for delay measurements
within their corresponding achromatic phases. The remaining nine LSST color curves are shown in Figure \ref{fig:color curves from LSST for different supernova model not useful}. The most promising
colors are rest-frame \textit{u-i}, \textit{g-r}, \textit{g-i}, and
\textit{r-i}. The colors \textit{u-g} and \textit{u-r} are also
encouraging but only if early features are captured. Therefore, to
target the most promising color curves, rest-frame filters \textit{u},
\textit{g}, \textit{r}, and \textit{i} are necessary, which lead to
three independent color curves. If one looks at the median color curves of the empirical SN model \texttt{SNEMO15} in Figure \ref{fig: SNEMO 15 color curves comparsion to models},
 \textit{g-r} and \textit{g-i} are almost featureless within the
 achromatic phase in comparison to the theoretical SNe models. Nevertheless the message of the
filters to target is not influenced since the other four promising
color curves include filters \textit{g}, \textit{r}, and \textit{i}
anyway.  

If one considers the median source redshift of the OM10 of $\sourcez =
0.77$, the \textit{r} band would be shifted from around
$\SI{6200}{\angstrom}$ to $\SI{11000}{\angstrom}$ and therefore mostly
not observed in the LSST bands. This means that from the six promising
color curves just rest-frame \textit{u-g} will be fully observed within the LSST
bands. Assuming the nearby iPTF16geu system ($\sourcez = 0.409$,
\cite{Goobar:2016uuf}), rest-frame \textit{u-g}, \textit{u-r}, and \textit{g-r}
would be observed fully but colors containing rest-frame \textit{i} band only
partly. This suggests that follow-up observations should also be
conducted in the infrared, but the results show the problem that even
though we find on average an achromatic phase of around three rest-frame
weeks for nearly all colors, more than half of them might not be
useful for time-delay measurements due to the lack of features in the
color curves. A case where redshifted color curves are investigated is
discussed in Section \ref{sec:Redshifted color curves}.

%\begin{figure*}
%\centering
%\includegraphics[scale=1]{color_curves}
%\caption{All rest-frame LSST color curves without microlensing for
%  four different SN Ia models in comparison to SN2011fe. The vertical
%  lines mark the mean duration of the achromatic phase by averaging
%  over 30 microlensing magnification maps (see Appendix \ref{sec:Appendix
%    achromatic phase in detail}).  Red frames indicate the color
%  curves that are promising for time-delay measurements (i.e., color
%  curves exhibiting features like extreme points or turning points
%  that are located within the achromatic phase).}
%\label{fig:color curves from LSST for different supernova model}
%\end{figure*}

\begin{figure*}
\centering
\includegraphics[width = 0.98 \textwidth]{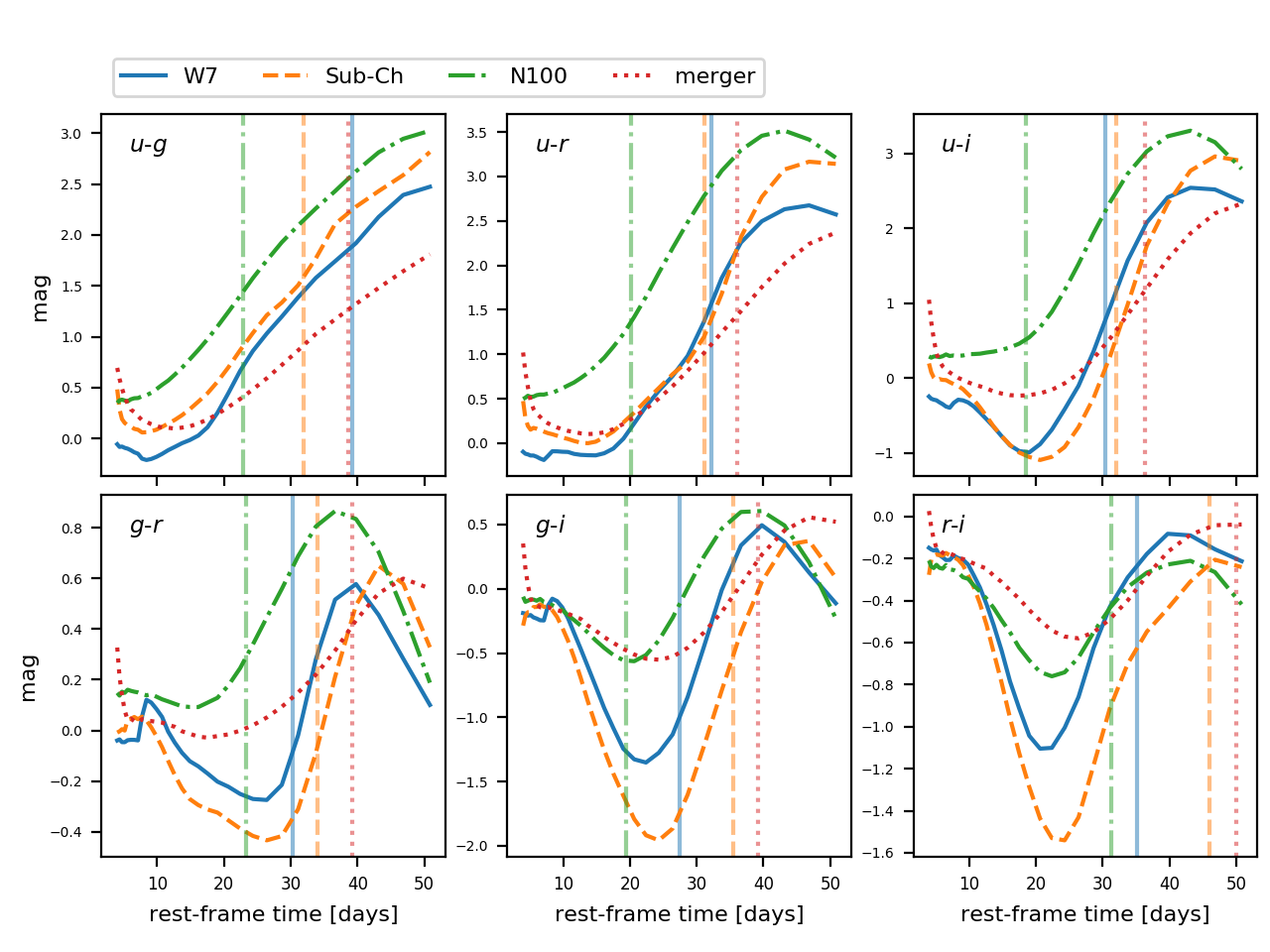}
\caption{Six rest-frame LSST color curves without microlensing for
  four different SN Ia models. The vertical
  lines mark the mean duration of the achromatic phase by averaging
  over 30 microlensing magnification maps (see Appendix \ref{sec:Appendix
    achromatic phase in detail}).  This figure contains only the color
  curves that are promising for time-delay measurements (i.e., color
  curves exhibiting features like extreme points or turning points
  that are located within the achromatic phase for at least three SN Ia models). For the remaining nine LSST color curves see Figure \ref{fig:color curves from LSST for different supernova model not useful}.}
\label{fig:color curves from LSST for different supernova model}
\end{figure*}

We emphasize that these results are based on averaging over the
investigated sample of magnification maps. In special cases, the
achromatic phase might be much shorter. The results are summarized in
Figure \ref{fig: achromatic phase as function of model, s and kappa
  gamma}, where the median and the 16th to 84th percentiles of the
achromatic-phase duration are shown. From the left-hand panel, we find
that the SN Ia models W7, \su \, and the merger yield on average a
comparable achromatic-phase duration, and the N100 model has a shorter
one. 
This might be related to the fact that the flux predicted by
the N100 model around maximum light is too red in comparison to
observations, which comes from an iron group element layer around a
Ni-56 core \citep{Sim:2013nna}, influencing the specific intensity
profiles and therefore also the duration of the achromatic
phase. 
However, given the uncertainties on the duration (shown in the
left-hand panel) and our tests of different criteria for computing the
achromatic-phase duration, we conclude that 
%Considering the 16th to 84th percentile this trend is not
%supported and testing different criteria of the achromatic phase also
%suggests that the median duration of the achromatic phase for
%different models is more similar. 
%To summarize 
we do not find a
significant dependency of the average achromatic phase on the SN
model. In addition, we find from the right-hand panel that strong lensing
images with a high macro-magnification $\mu =
((1-\kappa)^2-\gamma^2)^{-1}$ are influenced more by microlensing than
low magnification cases. Concerning the $s$ value, we find similar
durations of the achromatic phase for $s \le 0.5$ and an increased
duration if we go to smoother maps (middle panel). Depending on the
position of the lensed images, an appropriate $s$ value can be chosen
\citep{2011MNRAS.415.2215B,2014MNRAS.439.2494O,2015ApJ...799..149J}. Overall
we can say that combinations of $\kappa, \gamma$ and $s$ producing
smoother microlensing maps yield a longer achromatic phase.

The spread of the 1$\sigma$ range in Figure \ref{fig: achromatic phase
  as function of model, s and kappa gamma} is large because there is
quite some variation between different colors and microlensing maps,
which makes it hard to give a general recipe for using color curves
for time-delay measurements. Still, the median values in Figure \ref{fig: achromatic phase as function of model, s and kappa gamma} are around three rest-frame weeks or longer and therefore follow-up resources for LSNe Ia should be allocated independent of the lensing parameters $\kappa, \gamma$ and $s$.

\begin{figure*}[htbp]
\centering
\subfigure{\includegraphics[trim=10 10 10 10,clip,width=.32\textwidth]{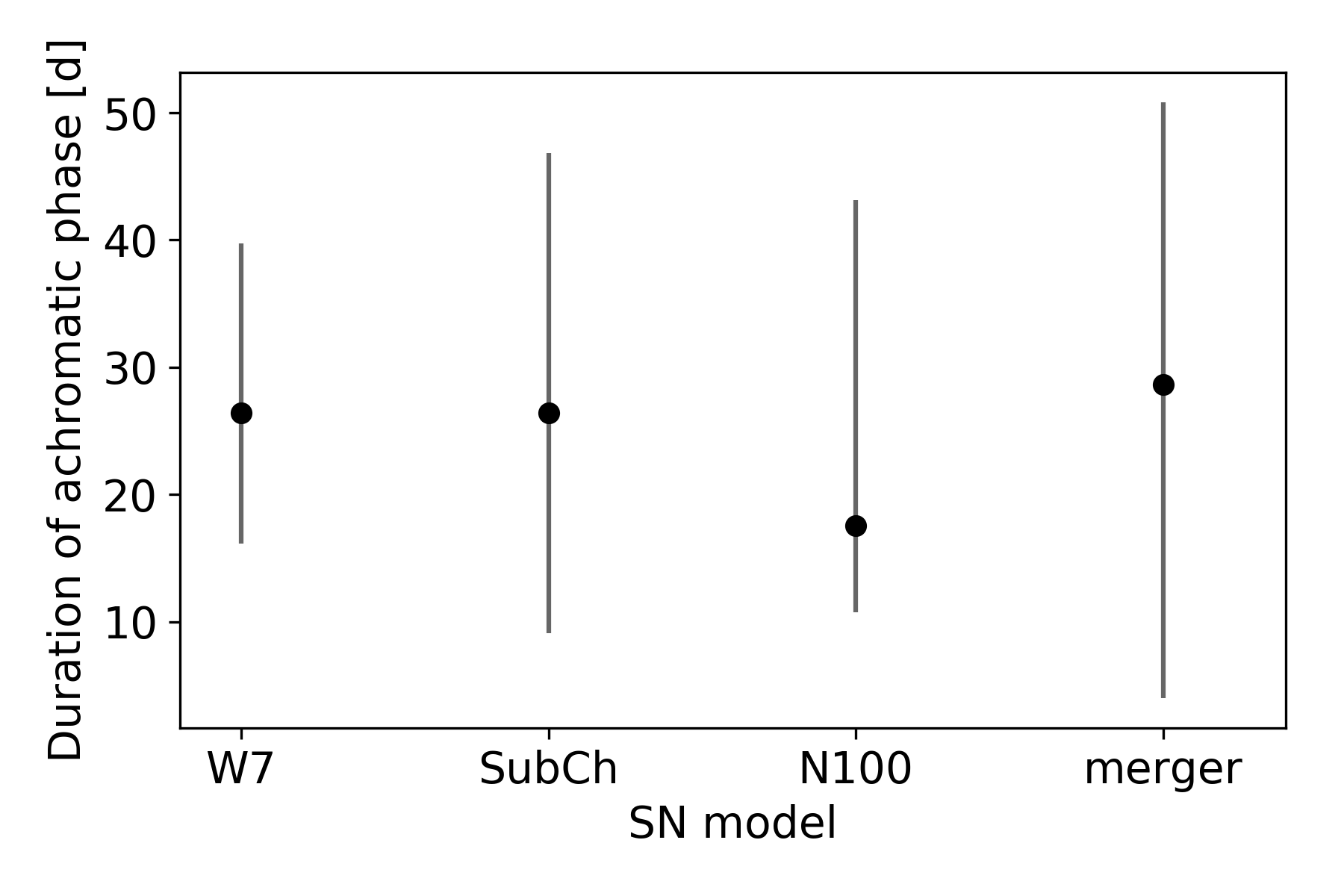}}
\subfigure{\includegraphics[trim=10 10 10 10,clip,width=.32\textwidth]{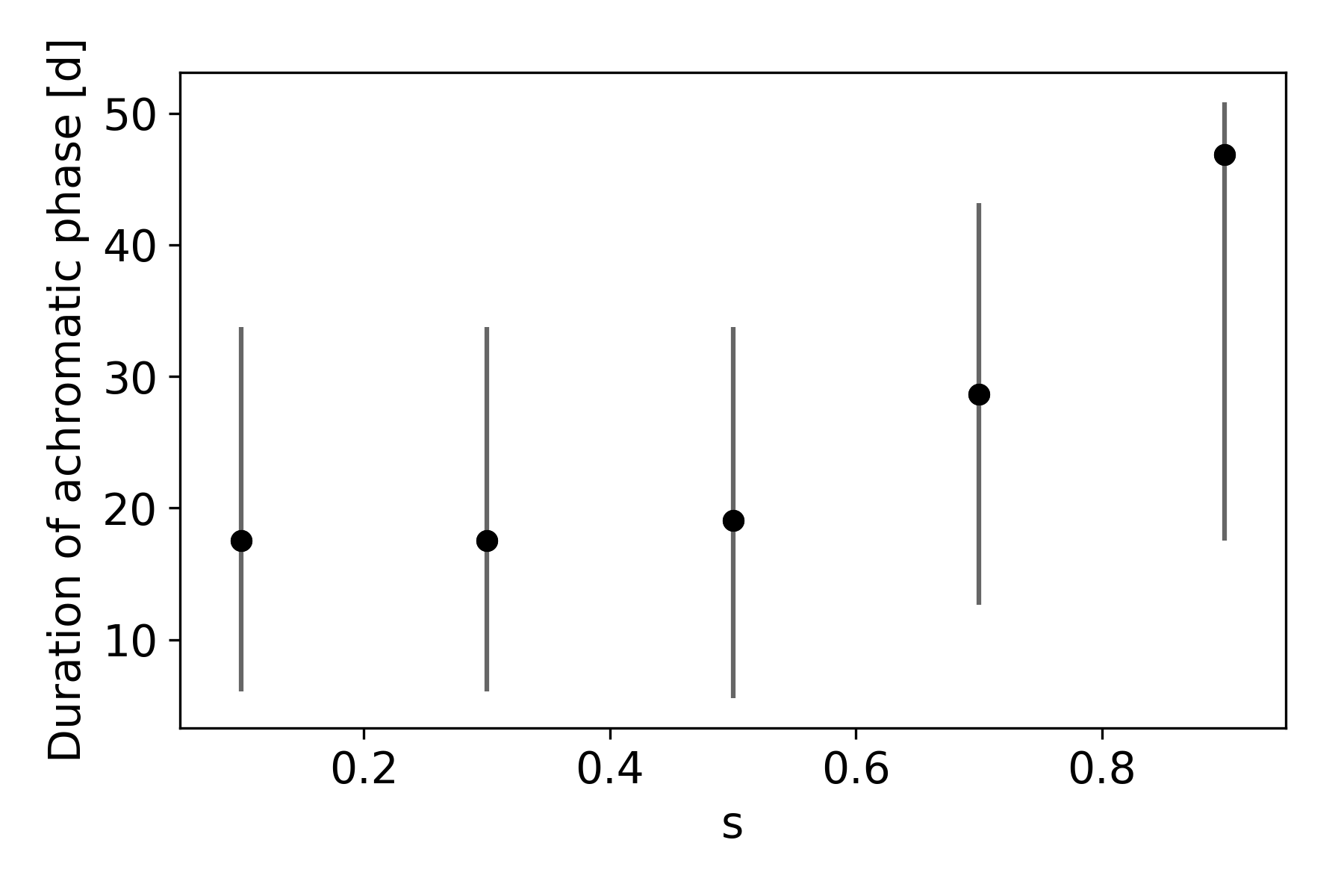}}
\subfigure{\includegraphics[trim=10 10 10 10,clip,width=.32\textwidth]{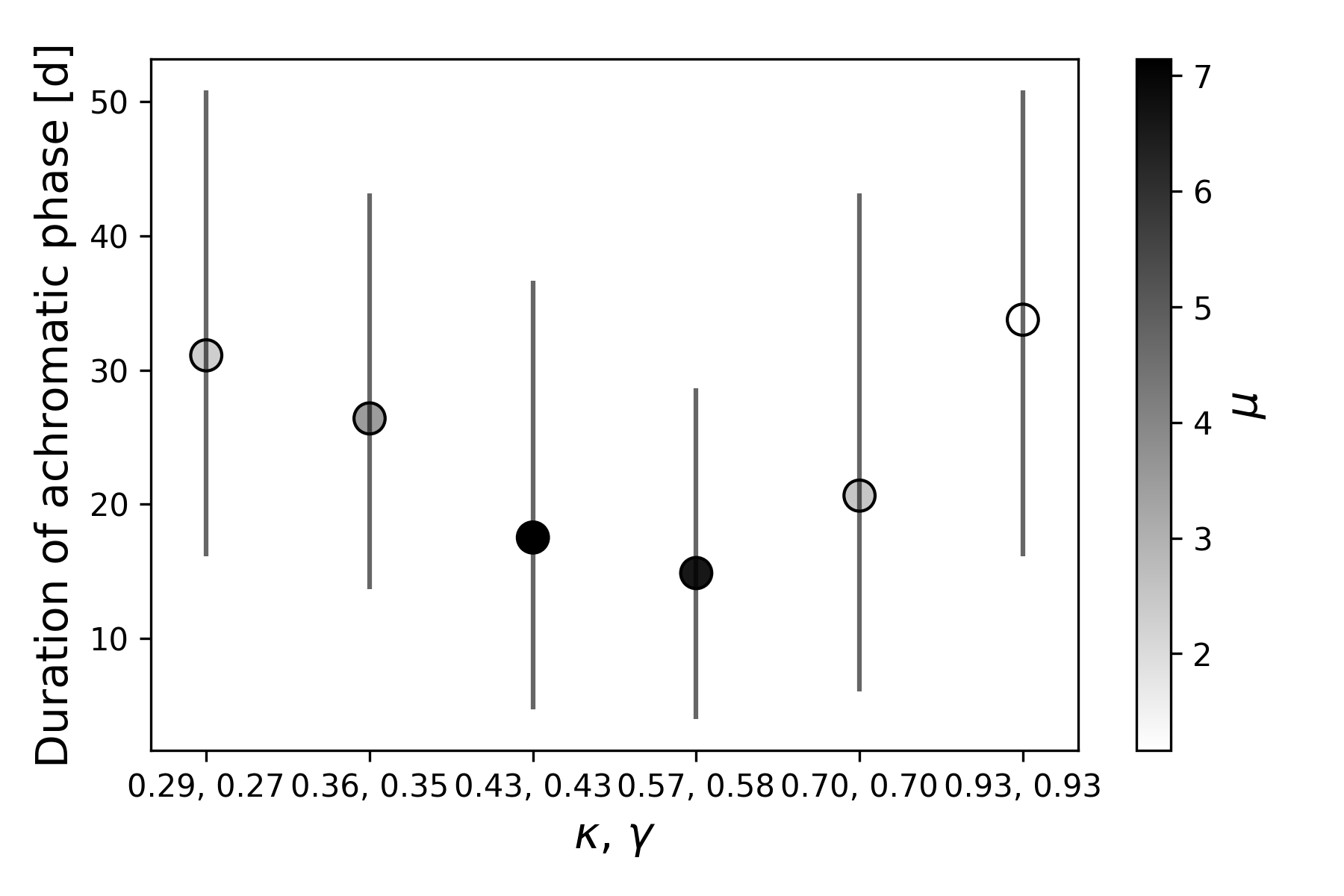}}
\caption{Duration of the achromatic phase in rest-frame days for
  different SN Ia models (left-hand panel), different smooth matter
  fractions $s$ (middle panel), and image configurations from strong
  lensing with their respective magnification factors $\mu$ shown in
  the color bar (right-hand panel). For these plots, a sample of four
  models, six different $\kappa$ and $\gamma$ values, five different $s$
  values, and 15 color curves has been investigated where we average
  over all parameters not shown on the $x$-axis. The dots correspond
  to the median, and the vertical bars indicate the range from the
  16th to 84th percentiles. Due to limits in computing time, we only
  investigated the achromatic phase up to rest-frame day 50.8;
  therefore, for cases close to that limit, the presented result are a
  lower limit on the achromatic phase. While the median achromatic
  phase is typically $\gtrsim$20 rest-frame days, the spread due to
  different microlensing maps and color curves is quite large and the
  results can be seen in detail in Appendix \ref{sec:Appendix
    achromatic phase in detail}.}
\label{fig: achromatic phase as function of model, s and kappa gamma}
\end{figure*}

\subsection{Scale of magnification map}
\label{sec: Redshift Dependency}
To probe the dependency on the scale of the microlensing map, namely
$\Rein$, we investigated for $\kappa,\gamma = (0.36, 0.35)$ and
$\kappa,\gamma = (0.70, 0.70)$ a range of redshifts, $(\sourcez,
\lensz) = \{(0.77,0.32), (0.55,0.16), (0.99,0.48)$, and $(0.99,0.16)$\}. The
first pair corresponds to the median values from the OM10 and the
second and third pair are the 16th and 84th percentile separately taken
for $\sourcez$ and $\lensz$. The fourth pair is the 84th percentile
for $\sourcez$ and 16th percentile of $\lensz$ to increase the variety
of different $\Rein$. The results are summarized in Figure
\ref{fig:Duration of achromatic phase as function of Einstein
  radius}. We find a very slight trend that with larger $\Rein$ the duration of
the achromatic phase becomes longer. Given the large uncertainties, we cannot report this trend to be significant, even though it would be plausible because with larger $\Rein$ 
the physical size of the magnification maps increases, which
causes SNe Ia to appear smaller in these maps, and for example events where micro
caustics are crossed will be less likely. Nevertheless, Figure \ref{fig:Duration of achromatic phase as function of Einstein
  radius} suggests that, if present at all, this effect is only minor.

\begin{figure}
\includegraphics[scale = 0.5]{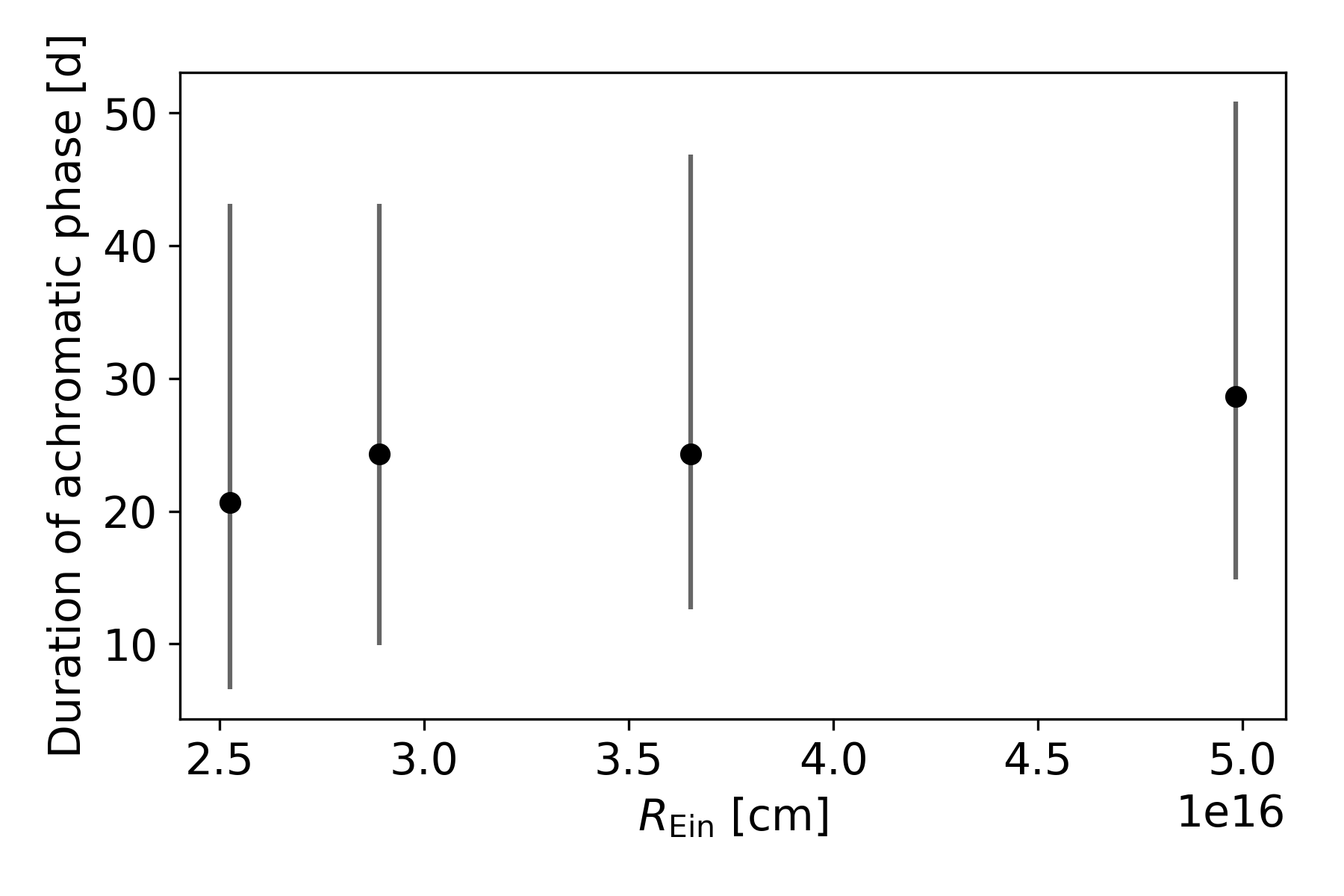}
\caption{Duration of achromatic phase as a function of $\Rein$, the
  scale of variations in the microlensing map. The dots correspond to
  the median and the vertical bars indicate the range from the 16th to
  84th percentile for the sample of four models, two different $\kappa$ and
  $\gamma$ pairs, five different $s$ values, and 15 color curves that have
  been investigated, where we average over all parameters not shown on
  the $x$-axis. From left to
  right, $\Rein$ corresponds to $(\sourcez, \lensz) = (0.99, 0.48),
  (0.77, 0.32), (0.55, 0.16), (0.99, 0.16)$.}
\label{fig:Duration of achromatic phase as function of Einstein radius}
\end{figure}

\subsection{Asymmetric merger model}
\label{sec: Asymetric Merger Model}

In this section, viewing angle effects are investigated for the
asymmetric merger model.  We compare the spherically symmetric
approach, which takes into account photon packets leaving the SN Ia
ejecta in any direction and averages over them to get the 1D
dependency on the impact parameter $p$, to six cases where only photons
from one half of the ejecta are taken into account, for example, just photons
that leave the ejecta at a positive $x$-coordinate, which we label as
$x>0$. For this subset we also calculate the 1D impact parameter, but
just averaging over photons leaving the ejecta at $x>0$.  The other
cases that we investigate are $x<0$, $y>0$, $y<0$, $z>0$ and $z<0$,
and the results are shown in Figure \ref{fig:color curves from LSST
  for asymmetric merger models}. We find that viewing angle effects
influence shapes of color curves and also the duration of the
achromatic phase, but only slightly and therefore useful color curves
(marked by red frames) are the same as those pointed out in Section
\ref{sec: SNe Ia models achromatic phase investigation}.

\begin{figure*}
\centering
\includegraphics[scale=1.2]{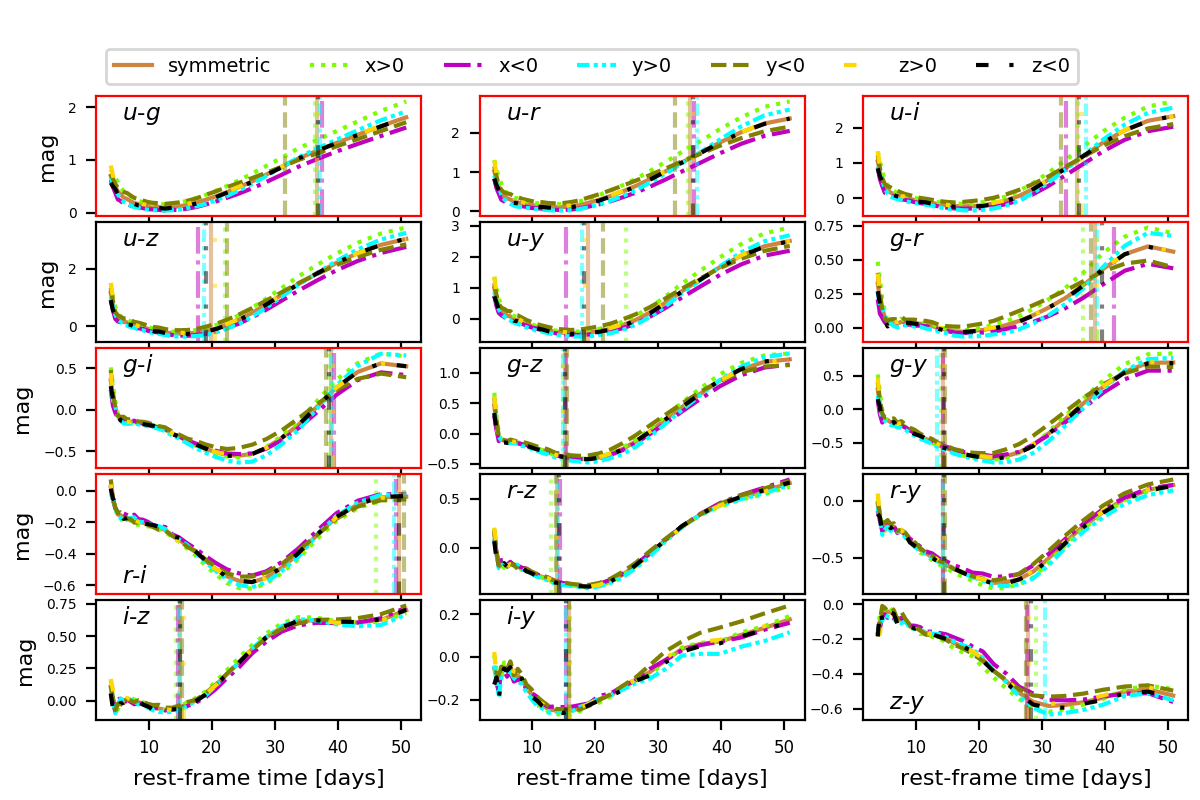}
\caption{All rest-frame LSST color curves without microlensing for the
  merger model assuming spherical symmetry and six asymmetric versions
  of the merger model. The vertical lines indicate the duration of the
  achromatic phase and red frames mark color curves that are
  promising for time-delay measurements.  The similarities between the
  symmetric case and the asymmetric cases show that viewing angle
  effects do not influence our conclusions.}
\label{fig:color curves from LSST for asymmetric merger models}
\end{figure*}

%\begin{figure*}[htbp]
%\centering
%\subfigure{\includegraphics[trim=30 17 22 16,clip,width=.32\textwidth]{figures/achromatic_phase_min.png}}
%\subfigure{\includegraphics[trim=30 17 22 16,clip,width=.32\textwidth]{figures/achromatic_phase_mean.png}}
%\subfigure{\includegraphics[trim=30 17 22 16,clip,width=.32\textwidth]{figures/achromatic_phase_max.png}}
%\caption{Duration of achromatic phase for color curves from neighboring filters n (\textit{u-g}, \textit{g-r}, \textit{r-i}, \textit{i-z}, \textit{z-y}), next to neighboring n+1 (\textit{u-r},\textit{g-i}, \textit{r-z}, \textit{z-y}) and so on. Except for the N100 model we see that neighboring filters have the longest achromatic phase.}
%\label{fig: achromatic phase of neighboring filter}
%\end{figure*}

\subsection{Redshifted color curves}
\label{sec:Redshifted color curves}

Section \ref{sec: SNe Ia models achromatic phase investigation}
indicates that most useful color curves (curves with features for
measuring time delays within the achromatic phase) in the SN
rest-frame will be shifted to the infrared regime for typical
redshifts expected for LSNe Ia. This section investigates if useful
color curves can still be found in \textit{ugrizy} coming from the
rest-frame ultraviolet (UV), taking into account typical redshifts of
SNe Ia. For this, we consider a set of representative redshifts:
$\sourcez, \lensz = \{(0.55,0.16), (0.77, 0.32),$ and $(0.99, 0.48)$\}.

\begin{figure*}
\centering
\includegraphics[scale=1.2]{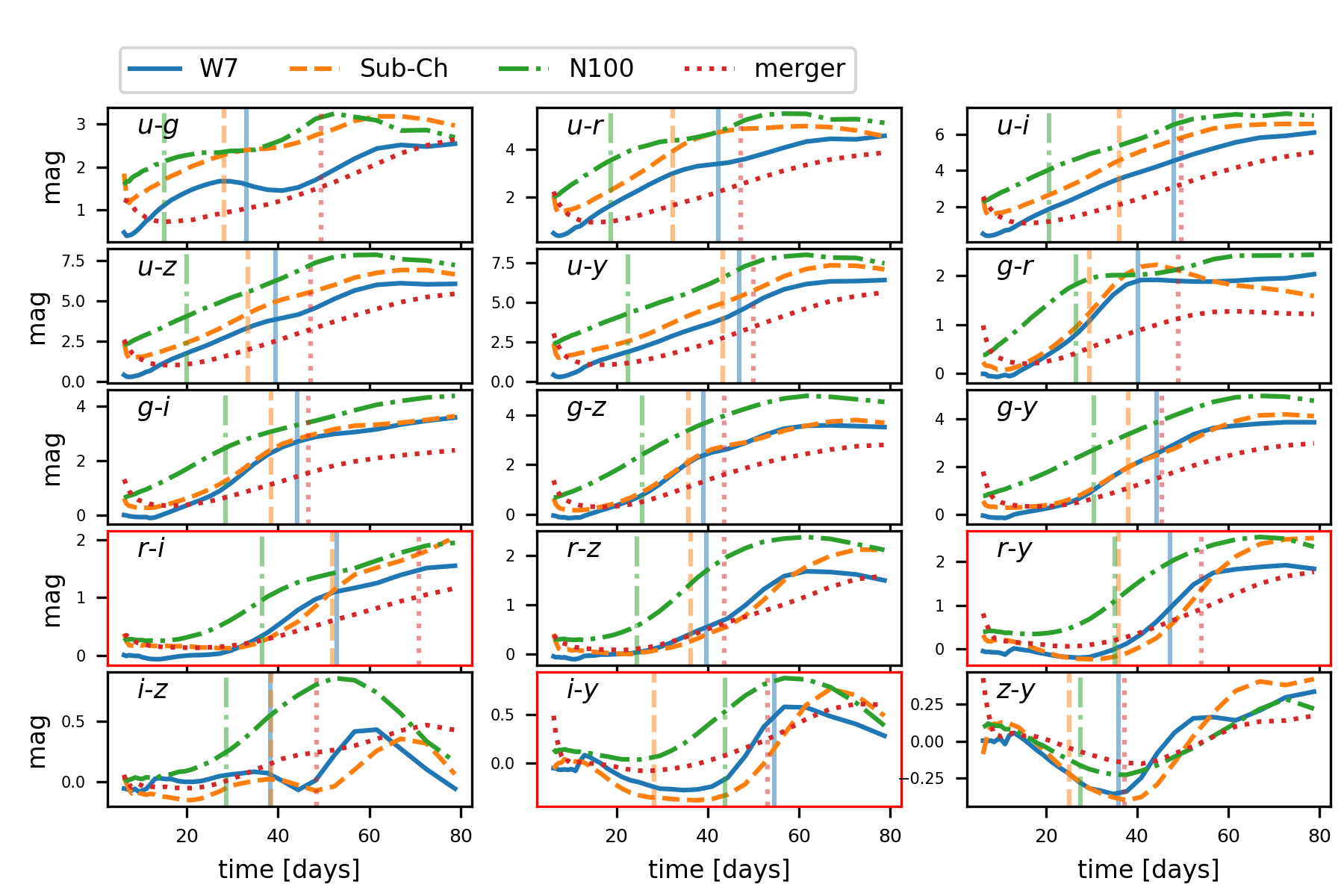}
\caption{All redshifted ($\sourcez = 0.55$) LSST color curves without
  microlensing for four different SN Ia models. The vertical lines mark
  the duration of the achromatic phase and red frames indicate color
  curves that are promising for time-delay measurements, in other words, color
  curves exhibiting features like extreme points or turning points
  that are located within the achromatic phase for at least three SNe Ia models.}
\label{fig:color curves redshifted by 0.55}
\end{figure*}

The results for $\sourcez = 0.55$ are shown in Figure \ref{fig:color
  curves redshifted by 0.55}. In comparison to the rest-frame bands,
we see from Figure \ref{fig:LSST filters transmission functions} that
the \textit{u} band will be observed in \textit{r}, the \textit{g}
band in \textit{i}, the \textit{r} band in \textit{y}, and the rest
will be in the infrared regime not covered by LSST
filters. Accordingly, we find three useful rest-frame color curves
from Figure \ref{fig:color curves from LSST for different supernova
  model} in Figure \ref{fig:color curves redshifted by 0.55}, namely
\textit{u-g} is redshifted roughly to \textit{r-i}, \textit{u-r} to
\textit{r-y}, and \textit{g-r} to \textit{i-y}, although only two of
the color curves are independent. Unfortunately none of the color
curves coming from the rest-frame UV show strong features within the
achromatic phase in at least three models that would be promising for
time-delay measurements. The cases for $ \sourcez = 0.77$ and
$\sourcez = 0.99$ are presented in Appendix \ref{sec:Appendix
  Redshifted color curves}. For $ \sourcez = 0.77$ we find two useful
independent color curves for time-delay measurement, namely
\textit{i-z} and \textit{z-y}, but at $\sourcez = 0.99$ there is only
\textit{z-y} remaining in the observed \textit{u} through \textit{y}
bands. For some color curves like \textit{g-i} for $\sourcez = 0.55$
or \textit{r-z} for $\sourcez = 0.77$ one might argue that they are
also useful if early features are captured but this is because they
still contain a substantial amount of rest-frame
\textit{u-g}. Nevertheless we see the trend that with higher redshifts
useful color curves are shifted to bands covering higher wavelengths
but there are no useful color curves coming from the rest-frame
UV. Therefore, the number of useful color curves observed in \textit{u}
through \textit{y} bands decreases with higher redshift. As discussed
by \cite{Suyu:2020opl}, the exact spectral shapes particularly of the
rest-frame UV spectra depend on various approximations used in the
radiative transfer calculations, such as metallicity of the progenitor or
the number of ionization states \citep[e.g.,][]{Lucy:1999, Lentz+2000,
  Kromer:2009ce, Walker+2012, Dessart+2014, Kromer+2016,
  Noebauer:2017vsf}, and therefore the shapes of color curves in the
rest-frame UV are more uncertain. With future more detailed radiative
transfer calculations, one might find useful color curves also in the
rest-frame UV but chances are low since only rest-frame UV colors of
the merger model show features within the achromatic phase, and all
other models do not (see Appendix \ref{sec:Appendix Redshifted color
  curves}).

Nevertheless these results suggest we need to follow up more in the infrared
regime that corresponds to the promising rest-frame color curves shown
in Figure \ref{fig:color curves from LSST for different supernova
  model}, especially for $\sourcez \gtrsim 0.6$. Apparent magnitudes
for the six LSST filters as well as three infrared bands (\textit{J, H} and
\textit{K}) are shown in Figure \ref{fig:light curves
  redshifted}. From this we find that in the \textit{u} and \textit{K}
bands the light curves are too faint but follow-up in all other bands
seems reasonable. \cite{Huber:2019ljb} have investigated light curves
for time-delay measurement and found that the combination of
\textit{g, r} and \textit{i} performs a few percent better than
\textit{r, i} and \textit{z}. The reason for this is the assumed
LSST-like 5$\sigma$ depth, which is nearly two magnitudes shallower in
the \textit{z} band in comparison to the \textit{g} band (in single
visits). In the context of this work, \textit{riz} might be chosen
over \textit{gri} for having more useful color curves. Nevertheless,
three bands are not ideal since this would result in just two
independent color curves. The more filters used for follow-up, the
higher the chances are to get promising color curves. A set of six
filters with \textit{r, i, z, y, J,} and \textit{H} seems most
promising as this would include all useful rest-frame LSST color
curves shown in Figure \ref{fig:color curves from LSST for different
  supernova model} for typical source redshifts. If resources for a
further band are available, then also the \textit{g} band can be used.
Covering this range of bands with a single follow-up telescope might
be challenging. If only an optical or only an infrared facility is
available, then we recommend to observe in the redder parts of the
optical coverage (\textit{rizy} bands) since these bands would yield
better quality light curves given that LSNe Ia are bright there and
observational uncertainties are lower than in the infrared
regime. Only for high redshift cases ($\sourcez \gtrsim 0.9$) it would
be worth to prefer the near-infrared over the optical range.

\begin{figure}
\includegraphics[scale=0.6]{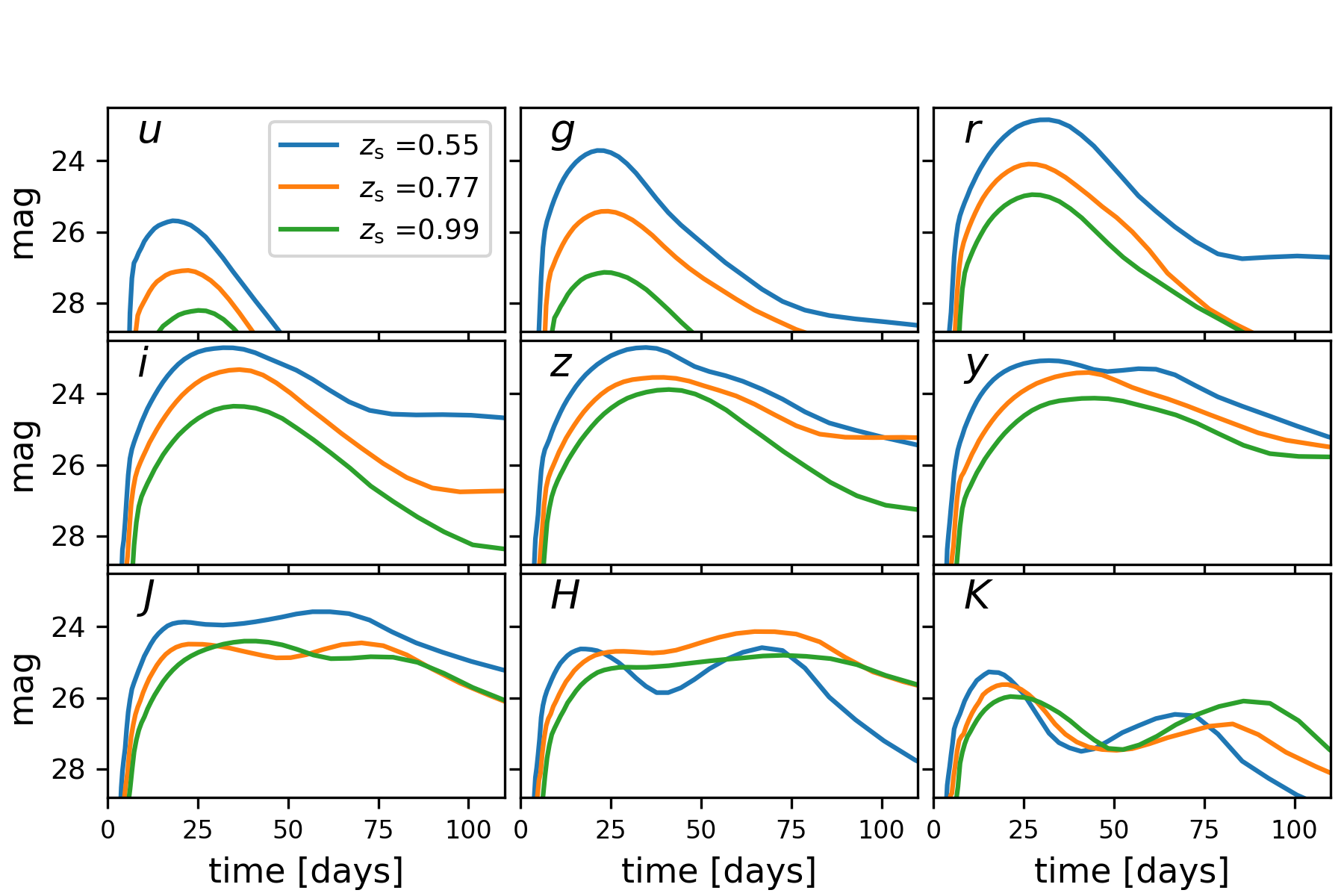}
\caption{Redshifted observed light curves for nine different filters and
  three different redshifts assuming the \su \, model. Light curves are
  too faint in the \textit{u} and \textit{K} bands but all other bands
  are potential candidates for follow-up observations.}
\label{fig:light curves redshifted}
\end{figure}

\section{Discussion and summary}
\label{sec: Discussion and summary}
According to \cite{Goldstein:2017bny}, the achromatic phase lasts for
three rest-frame weeks for the W7 model \citep{1984:Nomoto}, which
means that in this time frame, color curves are nearly independent of
microlensing and therefore promising for time-delay measurements. In
this work, we investigate in addition to the W7 model, a \su \,
model \citep{Sim:2010}, a merger model \citep{Pakmor:2012}, and a 3D
delayed detonation model \citep[model N100 of][]{Seitenzahl:2013}. Our
results are in good agreement with \cite{Goldstein:2017bny} leading on
average to an achromatic phase around three rest-frame weeks or
longer. Furthermore, we do not find a significant model dependency and
also asymmetries in the merger model do not have a strong influence on
the duration of the achromatic phase. While this sounds very promising
for time-delay measurements, there are also downsides to report. From
the 15 rest-frame LSST color curves, only six show promising features
for time-delay measurements and just three of them are
independent. These color curves contain combinations of the rest-frame
filters \textit{u, g, r,} and \textit{i}. To observe these for typical
LSN Ia redshifts, follow-up from bands \textit{r} to \textit{H} is
necessary. In an ideal follow-up scenario, one would observe in bands
\textit{r, i, z, y, J, H,} and optionally also \textit{g}. The bare
minimum should cover \textit{r, i,} and \textit{z}. Observations just
in three filters would make time-delay measurements from color curves
hard, but in \textit{riz} one can expect good quality light curves
that can also be used for time-delay measurements
\citep{Huber:2019ljb} although microlensing uncertainties are larger.

Even though the duration of the average achromatic phase is
around three rest-frame weeks or longer for most color curves, the
spread of the duration is quite large.  Depending on the configuration
in the microlensing map, a very short achromatic phase of just a few
days is also possible. Overall we find a longer achromatic phase for
smoother microlensing maps (high $s$ value) and image configurations with
lower magnification factors. The trend that combinations of $\sourcez, \lensz$,
which yield a larger Einstein radius $\Rein$, provide a longer achromatic phase is only very weak. Even though low $s$ values and high magnification cases provide a shorter duration of the achromatic phase, the median for these images is still around three rest-frame weeks. Therefore, lens and image properties of a LSNe Ia ($\sourcez$, $\lensz$, $\kappa$, $\gamma$ and $s$) can be mostly neglected when allocating follow-up observations, except $\sourcez$, which sets the filters to target.

This study provides general guidance on the observing filters to
follow up LSN Ia. In a real detection of a LSNe Ia where we have the
SN redshift measurement, we can further refine and optimize the
filters we have to employ to get promising color curves on a
case-by-case basis.

\begin{acknowledgements}
We thank M.~Oguri and P.~Marshall for the useful lens catalog  from  \cite{Oguri:2010}, and W.~Hillebrandt, S.~Blondin, D. A.~Goldstein for useful discussions.
We also would like to thank the anonymous referee for helpful comments, which strengthened this work.
SH and SHS thank the Max Planck Society for support through the Max
Planck Research Group for SHS. This project has received funding from
the European Research Council (ERC) under the European Union’s Horizon
2020 research and innovation programme
%SHS1103: revised to include COSMICLENS for Dominique
(LENSNOVA: grant agreement No 771776; COSMICLENS: grant
agreement No 787886).
%
%SHS1103: added
This research is supported in part by the Excellence Cluster ORIGINS which is funded by the Deutsche Forschungsgemeinschaft (DFG, German Research Foundation) under Germany's Excellence Strategy -- EXC-2094 -- 390783311.
UMN has been supported by the Transregional Collaborative Research
Center TRR33 ‘The Dark Universe’ of the Deutsche
Forschungsgemeinschaft.
JHHC acknowledges support from the Swiss National Science
Foundation and through European Research Council (ERC) under the European
Union's Horizon 2020 research and innovation programme (COSMICLENS:
grant agreement No 787866).
MK acknowledges support from
the Klaus Tschira Foundation. 
%SHS-waitMOconfirm% %
%SHS-waitMOconfirm% This work was supported in part by World Premier International
%SHS-waitMOconfirm% Research Center Initiative (WPI Initiative), MEXT, Japan, and JSPS
%SHS-waitMOconfirm% KAKENHI Grant Number JP15H05892 and JP18K03693.
%

\end{acknowledgements}

\bibliographystyle{aa}
%\bibliography{Lit}
\bibliography{Microlensing_SNeIa_models}

\clearpage
\appendix
\onecolumn
\section{Microlensing maps}
\label{sec: Microlensing maps}

\begin{figure*}[htbp]
\centering
\includegraphics[width=.7\textwidth]{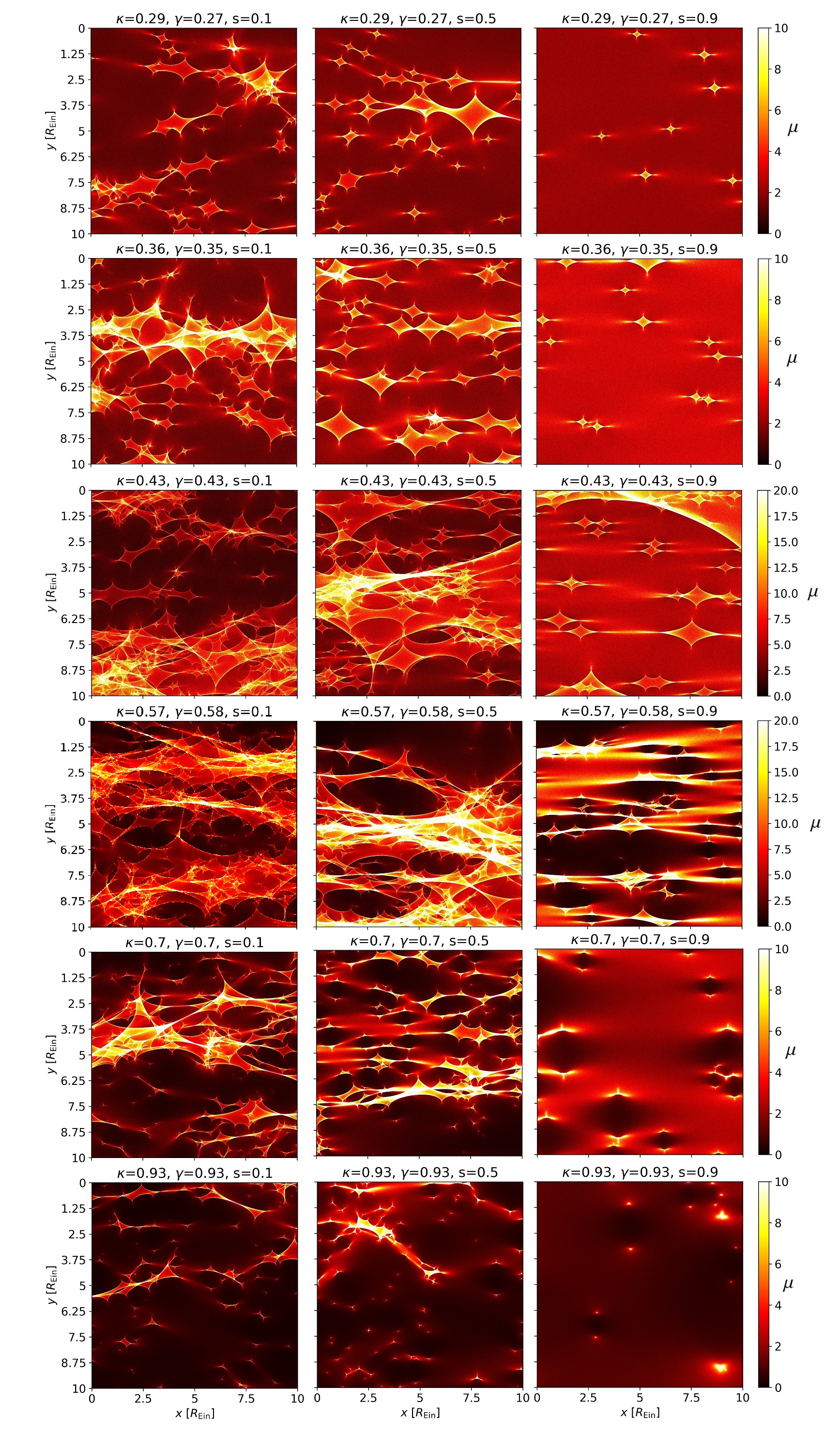}
\caption{Magnification maps for six different $\kappa$ and $\gamma$ values with smooth matter fraction $s = 0.1, 0.5,$ and $0.9$.}
\label{fig: microlensing maps different kappa, gammas same s}
\end{figure*}

\section{Achromatic phase in detail}

\label{sec:Appendix achromatic phase in detail}

\begin{table}[h]
\scalebox{0.91}{
\begin{tabular}{llllllllllllllllll}
\toprule
{} &            $\kappa$, $\gamma$ &    s &            u-g &            u-r &            u-i &            u-z &            u-y &            g-r &            g-i &            g-z &            g-y &            r-i &            r-z &            r-y &            i-z &            i-y &            z-y \\
\midrule
\multirow{30}{*}{W7}   &  \multirow{5}{*}{0.36 , 0.35} &  0.1 &           39.8 &           31.1 &           28.7 &           19.0 &           19.0 &           31.1 &           24.3 &           16.2 &           16.2 &           31.1 &           16.2 &           14.9 &           17.5 &           17.5 &           31.1 \\
                       &                               &  0.3 &           36.6 &           31.1 &           28.7 &           19.0 &           17.5 &           26.4 &           24.3 &           14.9 &           14.9 &           31.1 &           14.9 &           13.7 &           17.5 &           17.5 &           31.1 \\
                       &                               &  0.5 &           36.6 &           31.1 &           26.4 &           17.5 &           17.5 &           26.4 &           24.3 &           14.9 &           14.9 &           28.7 &           14.9 &           13.7 &           17.5 &           16.2 &           31.1 \\
                       &                               &  0.7 &           39.8 &           31.1 &           31.1 &           36.6 &           33.8 &           31.1 &           24.3 &           17.5 &           19.0 &           50.8 &           17.5 &           16.2 &           19.0 &           19.0 &           33.8 \\
                       &                               &  0.9 &           50.8 &           39.8 &           39.8 &           46.8 &           39.8 &           36.6 &           36.6 &           50.8 &           39.8 &           50.8 &           43.2 &           50.8 &           43.2 &           50.8 &           43.2 \\
                       \cline{3-18}
                       &  \multirow{5}{*}{0.29 , 0.27} &  0.1 &           39.8 &           31.1 &           28.7 &           33.8 &           31.1 &           31.1 &           24.3 &           16.2 &           17.5 &           31.1 &           16.2 &           14.9 &           19.0 &           17.5 &           31.1 \\
                       &                               &  0.3 &           39.8 &           31.1 &           28.7 &           33.8 &           31.1 &           31.1 &           24.3 &           16.2 &           17.5 &           46.8 &           16.2 &           16.2 &           17.5 &           17.5 &           33.8 \\
                       &                               &  0.5 &           39.8 &           31.1 &           28.7 &           19.0 &           19.0 &           31.1 &           24.3 &           16.2 &           16.2 &           31.1 &           16.2 &           14.9 &           17.5 &           17.5 &           31.1 \\
                       &                               &  0.7 &           46.8 &           33.8 &           33.8 &           39.8 &           36.6 &           33.8 &           31.1 &           39.8 &           36.6 &           50.8 &           43.2 &           22.4 &           24.3 &           22.4 &           39.8 \\
                       &                               &  0.9 &           50.8 &           50.8 &           50.8 &           50.8 &           50.8 &           50.8 &           50.8 &           50.8 &           50.8 &           50.8 &           50.8 &           50.8 &           50.8 &           50.8 &           50.8 \\
                       \cline{3-18}
                       &  \multirow{5}{*}{0.43 , 0.43} &  0.1 &           39.8 &           31.1 &           28.7 &           19.0 &           19.0 &           31.1 &           24.3 &           16.2 &           16.2 &           31.1 &           16.2 &           14.9 &           17.5 &           17.5 &           31.1 \\
                       &                               &  0.3 &           39.8 &           31.1 &           28.7 &           17.5 &           17.5 &           26.4 &           24.3 &           14.9 &           14.9 &           31.1 &           14.9 &            4.0 &           17.5 &           17.5 &           31.1 \\
                       &                               &  0.5 &           36.6 &           31.1 &           26.4 &           17.5 &           17.5 &           26.4 &           24.3 &           14.9 &           14.9 &           28.7 &           14.9 &           13.7 &           17.5 &           16.2 &           31.1 \\
                       &                               &  0.7 &           33.8 &           28.7 &           26.4 &           17.5 &           16.2 &           26.4 &           24.3 &           14.9 &           13.7 &           28.7 &           14.9 &            4.0 &           17.5 &           16.2 &           31.1 \\
                       &                               &  0.9 &           39.8 &           31.1 &           31.1 &           33.8 &           33.8 &           31.1 &           24.3 &           16.2 &           17.5 &           50.8 &           16.2 &           16.2 &           19.0 &           17.5 &           33.8 \\
                       \cline{3-18}
                       &  \multirow{5}{*}{0.70 , 0.70} &  0.1 &           33.8 &           26.4 &           24.3 &           16.2 &           14.9 &           24.3 &           22.4 &           13.7 &           13.7 &           24.3 &           13.7 &            4.0 &           16.2 &           13.7 &           28.7 \\
                       &                               &  0.3 &           33.8 &           26.4 &           24.3 &           16.2 &           16.2 &           24.3 &           22.4 &           14.9 &           13.7 &           26.4 &           14.9 &           13.7 &           17.5 &           14.9 &           28.7 \\
                       &                               &  0.5 &           31.1 &           26.4 &           24.3 &           14.9 &           13.7 &           24.3 &           22.4 &           12.6 &           11.6 &           17.5 &           12.6 &            4.0 &           16.2 &           13.7 &           22.4 \\
                       &                               &  0.7 &           33.8 &           31.1 &           28.7 &           33.8 &           31.1 &           28.7 &           24.3 &           17.5 &           17.5 &           28.7 &           17.5 &           16.2 &           19.0 &           19.0 &           31.1 \\
                       &                               &  0.9 &           50.8 &           39.8 &           43.2 &           46.8 &           43.2 &           39.8 &           39.8 &           50.8 &           43.2 &           50.8 &           43.2 &           50.8 &           43.2 &           50.8 &           43.2 \\
                       \cline{3-18}
                       &  \multirow{5}{*}{0.57 , 0.58} &  0.1 &           36.6 &           28.7 &           26.4 &           17.5 &           17.5 &           26.4 &           24.3 &           14.9 &           14.9 &           28.7 &           14.9 &            4.0 &           17.5 &           16.2 &           31.1 \\
                       &                               &  0.3 &           36.6 &           31.1 &           26.4 &           17.5 &           17.5 &           26.4 &           24.3 &           14.9 &           14.9 &           28.7 &           14.9 &            4.0 &           17.5 &           16.2 &           31.1 \\
                       &                               &  0.5 &           33.8 &           26.4 &           24.3 &           16.2 &           16.2 &           24.3 &           22.4 &           13.7 &           13.7 &           24.3 &           13.7 &            4.0 &           16.2 &           14.9 &           28.7 \\
                       &                               &  0.7 &           31.1 &           26.4 &           24.3 &           14.9 &           14.9 &           24.3 &           22.4 &           12.6 &            4.0 &           24.3 &            4.0 &            4.0 &           16.2 &           13.7 &           20.7 \\
                       &                               &  0.9 &           31.1 &           26.4 &           24.3 &           14.9 &           13.7 &           22.4 &           22.4 &           12.6 &           11.6 &           16.2 &           12.6 &            5.6 &           16.2 &           13.7 &           20.7 \\
                       \cline{3-18}
                       &  \multirow{5}{*}{0.93 , 0.93} &  0.1 &           33.8 &           28.7 &           26.4 &           19.0 &           17.5 &           26.4 &           24.3 &           16.2 &           16.2 &           28.7 &           16.2 &           14.9 &           17.5 &           17.5 &           31.1 \\
                       &                               &  0.3 &           33.8 &           28.7 &           26.4 &           19.0 &           17.5 &           26.4 &           24.3 &           16.2 &           16.2 &           28.7 &           16.2 &           14.9 &           17.5 &           17.5 &           31.1 \\
                       &                               &  0.5 &           43.2 &           33.8 &           31.1 &           36.6 &           33.8 &           33.8 &           28.7 &           20.7 &           33.8 &           50.8 &           19.0 &           19.0 &           20.7 &           20.7 &           36.6 \\
                       &                               &  0.7 &           50.8 &           39.8 &           39.8 &           46.8 &           39.8 &           36.6 &           36.6 &           50.8 &           39.8 &           50.8 &           43.2 &           50.8 &           43.2 &           50.8 &           43.2 \\
                       &                               &  0.9 &           50.8 &           50.8 &           50.8 &           50.8 &           50.8 &           50.8 &           50.8 &           50.8 &           50.8 &           50.8 &           50.8 &           50.8 &           50.8 &           50.8 &           50.8 \\
                       \cline{3-18}
                       &                          mean &      &  \textbf{39.2} &  \textbf{32.2} &  \textbf{30.4} &  \textbf{26.8} &  \textbf{25.3} &  \textbf{30.3} &  \textbf{27.4} &  \textbf{22.1} &  \textbf{21.2} &  \textbf{35.1} &  \textbf{21.1} &  \textbf{18.1} &  \textbf{22.6} &  \textbf{22.5} &  \textbf{33.1} \\
\midrule
\multirow{30}{*}{N100} &  \multirow{5}{*}{0.36 , 0.35} &  0.1 &           20.7 &           17.5 &           14.9 &           31.1 &           17.5 &           19.0 &           12.6 &           17.5 &           19.0 &           26.4 &           13.7 &            4.0 &            9.1 &           43.2 &           14.9 \\
                       &                               &  0.3 &           19.0 &           17.5 &           14.9 &           31.1 &           17.5 &           19.0 &           12.6 &           17.5 &           19.0 &           26.4 &           13.7 &            4.0 &            9.1 &           43.2 &           14.9 \\
                       &                               &  0.5 &           19.0 &           16.2 &           14.9 &           31.1 &           17.5 &           19.0 &           13.7 &           16.2 &           19.0 &           26.4 &           13.7 &           26.4 &            9.1 &           43.2 &           14.9 \\
                       &                               &  0.7 &           20.7 &           19.0 &           16.2 &           31.1 &           19.0 &           20.7 &           16.2 &           19.0 &           28.7 &           50.8 &           14.9 &           43.2 &           12.6 &           43.2 &           16.2 \\
                       &                               &  0.9 &           43.2 &           31.1 &           33.8 &           50.8 &           33.8 &           39.8 &           39.8 &           50.8 &           36.6 &           50.8 &           24.3 &           50.8 &           43.2 &           50.8 &           39.8 \\
                       \cline{3-18}
                       &  \multirow{5}{*}{0.29 , 0.27} &  0.1 &           20.7 &           17.5 &           16.2 &           31.1 &           19.0 &           20.7 &           14.9 &           17.5 &           20.7 &           26.4 &           14.9 &           43.2 &            9.9 &           43.2 &           14.9 \\
                       &                               &  0.3 &           20.7 &           17.5 &           16.2 &           31.1 &           19.0 &           20.7 &           16.2 &           19.0 &           26.4 &           28.7 &           14.9 &           43.2 &           11.6 &           43.2 &           16.2 \\
                       &                               &  0.5 &           20.7 &           17.5 &           14.9 &           31.1 &           19.0 &           19.0 &           14.9 &           17.5 &           20.7 &           26.4 &           13.7 &           43.2 &           10.7 &           43.2 &           14.9 \\
                       &                               &  0.7 &           31.1 &           26.4 &           28.7 &           36.6 &           31.1 &           33.8 &           36.6 &           43.2 &           36.6 &           50.8 &           20.7 &           46.8 &           20.7 &           50.8 &           20.7 \\
                       &                               &  0.9 &           50.8 &           46.8 &           46.8 &           43.2 &           46.8 &           50.8 &           50.8 &           50.8 &           50.8 &           50.8 &           50.8 &           50.8 &           50.8 &           50.8 &           46.8 \\
                       \cline{3-18}
                       &  \multirow{5}{*}{0.43 , 0.43} &  0.1 &           20.7 &           17.5 &           14.9 &           31.1 &           19.0 &           19.0 &           13.7 &           17.5 &           20.7 &           26.4 &            4.0 &            4.0 &            9.1 &           43.2 &           14.9 \\
                       &                               &  0.3 &           19.0 &           17.5 &           13.7 &           31.1 &           17.5 &           19.0 &           12.6 &           17.5 &           19.0 &           26.4 &            4.0 &            4.0 &            5.1 &           43.2 &           14.9 \\
                       &                               &  0.5 &           19.0 &           16.2 &           13.7 &           31.1 &           17.5 &           17.5 &           11.6 &           16.2 &           19.0 &           26.4 &            4.0 &            4.0 &            7.1 &            9.1 &           13.7 \\
                       &                               &  0.7 &           17.5 &           16.2 &           12.6 &           31.1 &           16.2 &           17.5 &           11.6 &            4.0 &            4.0 &           26.4 &            4.0 &            4.0 &            4.0 &            4.0 &           13.7 \\
                       &                               &  0.9 &           22.4 &           19.0 &           17.5 &           31.1 &           20.7 &           22.4 &           16.2 &           19.0 &           28.7 &           50.8 &           14.9 &           43.2 &           11.6 &           43.2 &           16.2 \\
                       \cline{3-18}
                       &  \multirow{5}{*}{0.70 , 0.70} &  0.1 &           14.9 &           13.7 &           10.7 &           28.7 &           14.9 &           16.2 &           10.7 &            4.0 &            4.0 &           13.7 &            4.0 &            4.0 &            4.0 &            4.0 &           12.6 \\
                       &                               &  0.3 &           14.9 &           13.7 &           11.6 &           26.4 &           14.9 &           16.2 &           11.6 &           13.7 &           17.5 &           13.7 &            5.1 &           24.3 &            7.1 &            7.7 &           12.6 \\
                       &                               &  0.5 &           13.7 &           12.6 &           10.7 &           24.3 &           14.9 &           16.2 &           10.7 &            4.0 &            4.0 &           12.6 &            4.0 &            4.0 &            4.0 &            4.0 &           12.6 \\
                       &                               &  0.7 &           19.0 &           17.5 &           14.9 &           28.7 &           17.5 &           19.0 &           16.2 &           17.5 &           20.7 &           26.4 &           14.9 &           43.2 &           11.6 &           43.2 &           14.9 \\
                       &                               &  0.9 &           36.6 &           31.1 &           33.8 &           43.2 &           33.8 &           39.8 &           39.8 &           50.8 &           39.8 &           50.8 &           39.8 &           50.8 &           39.8 &           50.8 &           39.8 \\
                       \cline{3-18}
                       &  \multirow{5}{*}{0.57 , 0.58} &  0.1 &           17.5 &           16.2 &           12.6 &           28.7 &           16.2 &           17.5 &           11.6 &           16.2 &           17.5 &           24.3 &            4.0 &            4.0 &            5.1 &            7.7 &           13.7 \\
                       &                               &  0.3 &           19.0 &           16.2 &           12.6 &           31.1 &           17.5 &           17.5 &           11.6 &           16.2 &           17.5 &           26.4 &            4.0 &            4.0 &            5.1 &            9.1 &           13.7 \\
                       &                               &  0.5 &           16.2 &           14.9 &           11.6 &           28.7 &           14.9 &           16.2 &           10.7 &            4.0 &            4.0 &           24.3 &            4.0 &            4.0 &            4.0 &            4.0 &           12.6 \\
                       &                               &  0.7 &           14.9 &           12.6 &            9.9 &           26.4 &           14.9 &           16.2 &            9.9 &            4.0 &            4.0 &           12.6 &            4.0 &            4.0 &            4.0 &            4.0 &           12.6 \\
                       &                               &  0.9 &           13.7 &           12.6 &            9.9 &           20.7 &           13.7 &           14.9 &           10.7 &            5.1 &            4.0 &           11.6 &            4.0 &            4.0 &            4.0 &            4.0 &           12.6 \\
                       \cline{3-18}
                       &  \multirow{5}{*}{0.93 , 0.93} &  0.1 &           19.0 &           16.2 &           13.7 &           28.7 &           17.5 &           19.0 &           13.7 &           16.2 &           19.0 &           26.4 &           13.7 &           26.4 &            9.9 &           43.2 &           14.9 \\
                       &                               &  0.3 &           17.5 &           16.2 &           13.7 &           28.7 &           16.2 &           19.0 &           13.7 &           16.2 &           19.0 &           24.3 &           13.7 &           26.4 &            9.9 &           39.8 &           14.9 \\
                       &                               &  0.5 &           22.4 &           20.7 &           20.7 &           31.1 &           22.4 &           24.3 &           31.1 &           20.7 &           31.1 &           50.8 &           17.5 &           43.2 &           14.9 &           43.2 &           17.5 \\
                       &                               &  0.7 &           31.1 &           26.4 &           28.7 &           36.6 &           31.1 &           36.6 &           36.6 &           50.8 &           36.6 &           50.8 &           24.3 &           50.8 &           36.6 &           50.8 &           24.3 \\
                       &                               &  0.9 &           50.8 &           50.8 &           50.8 &           50.8 &           50.8 &           50.8 &           50.8 &           50.8 &           50.8 &           50.8 &           50.8 &           50.8 &           50.8 &           50.8 &           50.8 \\
                       \cline{3-18}
                       &                          mean &      &  \textbf{22.9} &  \textbf{20.2} &  \textbf{18.5} &  \textbf{32.2} &  \textbf{21.4} &  \textbf{23.2} &  \textbf{19.5} &  \textbf{21.1} &  \textbf{22.0} &  \textbf{31.3} &  \textbf{14.5} &  \textbf{25.3} &  \textbf{14.5} &  \textbf{32.1} &  \textbf{18.9} \\
\midrule
\bottomrule
\end{tabular}
}
\caption{Duration of the achromatic phase $t_\mathrm{achro}$ in rest-frame days for all 15 LSST color curves for the W7 and N100 model and 30 different microlensing magnification maps ($\kappa, \gamma,$ and $s$) as in Section \ref{sec: SNe Ia models achromatic phase investigation}, where for each map 10000 random positions are drawn. The mean values are plotted in Figure \ref{fig:color curves from LSST for different supernova model}.}
\label{tab: duration of achromatic phase w7 and n100}
\end{table}

\begin{table}
\begin{tabular}{llllllllllllllllll}
\toprule
{} &            $\kappa$, $\gamma$ &    s &            u-g &            u-r &            u-i &            u-z &            u-y &            g-r &            g-i &            g-z &            g-y &            r-i &            r-z &            r-y &            i-z &            i-y &            z-y \\
\midrule
\multirow{30}{*}{\su}  &  \multirow{5}{*}{0.36 , 0.35} &  0.1 &           31.1 &           31.1 &           31.1 &           17.5 &           14.9 &           36.6 &           36.6 &           12.6 &           10.7 &           50.8 &           14.9 &            6.0 &           14.9 &            6.6 &           26.4 \\
                         &                               &  0.3 &           31.1 &           28.7 &           31.1 &           16.2 &           13.7 &           36.6 &           36.6 &           11.6 &            9.9 &           50.8 &           14.9 &            6.0 &           14.9 &            6.0 &           24.3 \\
                         &                               &  0.5 &           28.7 &           28.7 &           28.7 &           16.2 &           12.6 &           36.6 &           33.8 &           11.6 &            9.9 &           50.8 &           14.9 &            6.6 &           14.9 &            6.6 &           24.3 \\
                         &                               &  0.7 &           31.1 &           31.1 &           31.1 &           36.6 &           31.1 &           36.6 &           36.6 &           14.9 &           13.7 &           50.8 &           16.2 &           14.9 &           17.5 &           12.6 &           26.4 \\
                         &                               &  0.9 &           46.8 &           39.8 &           43.2 &           50.8 &           43.2 &           46.8 &           46.8 &           50.8 &           46.8 &           50.8 &           46.8 &           50.8 &           46.8 &           50.8 &           46.8 \\
                         \cline{3-18}
                         &  \multirow{5}{*}{0.29 , 0.27} &  0.1 &           31.1 &           31.1 &           31.1 &           19.0 &           17.5 &           36.6 &           36.6 &           12.6 &           11.6 &           50.8 &           16.2 &            7.7 &           16.2 &            7.7 &           26.4 \\
                         &                               &  0.3 &           31.1 &           31.1 &           31.1 &           36.6 &           31.1 &           36.6 &           36.6 &           13.7 &           12.6 &           50.8 &           16.2 &           11.6 &           17.5 &           10.7 &           26.4 \\
                         &                               &  0.5 &           31.1 &           31.1 &           31.1 &           20.7 &           17.5 &           36.6 &           36.6 &           12.6 &           11.6 &           50.8 &           16.2 &            9.9 &           17.5 &            9.1 &           26.4 \\
                         &                               &  0.7 &           33.8 &           33.8 &           36.6 &           43.2 &           36.6 &           43.2 &           43.2 &           50.8 &           43.2 &           50.8 &           43.2 &           50.8 &           39.8 &           50.8 &           39.8 \\
                         &                               &  0.9 &           50.8 &           50.8 &           50.8 &           50.8 &           50.8 &           50.8 &           50.8 &           50.8 &           50.8 &           50.8 &           50.8 &           50.8 &           50.8 &           50.8 &           50.8 \\
                         \cline{3-18}
                         &  \multirow{5}{*}{0.43 , 0.43} &  0.1 &           31.1 &           31.1 &           31.1 &           17.5 &           14.9 &           36.6 &           36.6 &           11.6 &            9.9 &           50.8 &           14.9 &            6.0 &           14.9 &            6.0 &           26.4 \\
                         &                               &  0.3 &           31.1 &           31.1 &           31.1 &           16.2 &           12.6 &           36.6 &           36.6 &           11.6 &            9.9 &           50.8 &            4.0 &            5.6 &            6.0 &            6.0 &           24.3 \\
                         &                               &  0.5 &           28.7 &           28.7 &           28.7 &           16.2 &           12.6 &           36.6 &           33.8 &           11.6 &            9.1 &           50.8 &            4.0 &            5.6 &            6.0 &            6.0 &           24.3 \\
                         &                               &  0.7 &           28.7 &           28.7 &           28.7 &           14.9 &            9.9 &           36.6 &           33.8 &            9.1 &            8.4 &           46.8 &            4.0 &            4.4 &            4.0 &            5.6 &           20.7 \\
                         &                               &  0.9 &           31.1 &           31.1 &           31.1 &           36.6 &           31.1 &           36.6 &           36.6 &           13.7 &           12.6 &           50.8 &           16.2 &           12.6 &           17.5 &           11.6 &           26.4 \\
                         \cline{3-18}
                         &  \multirow{5}{*}{0.70 , 0.70} &  0.1 &           26.4 &           26.4 &           26.4 &           12.6 &            8.4 &           12.6 &           28.7 &            6.0 &            6.6 &           28.7 &            4.0 &            4.4 &            4.0 &            4.4 &           19.0 \\
                         &                               &  0.3 &           26.4 &           28.7 &           28.7 &           14.9 &           10.7 &           13.7 &           31.1 &            9.9 &            9.1 &           31.1 &            6.0 &            6.0 &            6.6 &            6.0 &           20.7 \\
                         &                               &  0.5 &           26.4 &           26.4 &           26.4 &           11.6 &            8.4 &           12.6 &           16.2 &            6.6 &            7.7 &           26.4 &            4.0 &            4.4 &            4.0 &            5.6 &           17.5 \\
                         &                               &  0.7 &           28.7 &           28.7 &           28.7 &           33.8 &           31.1 &           36.6 &           36.6 &           14.9 &           14.9 &           46.8 &           17.5 &           14.9 &           17.5 &           14.9 &           26.4 \\
                         &                               &  0.9 &           39.8 &           36.6 &           39.8 &           50.8 &           39.8 &           46.8 &           46.8 &           50.8 &           46.8 &           50.8 &           46.8 &           50.8 &           46.8 &           50.8 &           43.2 \\
                         \cline{3-18}
                         &  \multirow{5}{*}{0.57 , 0.58} &  0.1 &           28.7 &           28.7 &           28.7 &           16.2 &           11.6 &           36.6 &           33.8 &            9.9 &            8.4 &           46.8 &            4.0 &            4.4 &            6.0 &            5.6 &           22.4 \\
                         &                               &  0.3 &           28.7 &           28.7 &           28.7 &           16.2 &           11.6 &           36.6 &           33.8 &           11.6 &            9.1 &           46.8 &            4.0 &            5.6 &            6.0 &            6.0 &           24.3 \\
                         &                               &  0.5 &           28.7 &           28.7 &           28.7 &           12.6 &            9.1 &           13.7 &           31.1 &            8.4 &            7.7 &           46.8 &            4.0 &            4.4 &            4.0 &            5.6 &           19.0 \\
                         &                               &  0.7 &           26.4 &           26.4 &           26.4 &           11.6 &            7.7 &           12.6 &           16.2 &            5.6 &            6.0 &           26.4 &            4.0 &            4.4 &            4.0 &            4.4 &           17.5 \\
                         &                               &  0.9 &           24.3 &           13.7 &           26.4 &           11.6 &            9.1 &           12.6 &           16.2 &            7.7 &            7.7 &           26.4 &            4.0 &            5.6 &            5.6 &            5.6 &           17.5 \\
                         \cline{3-18}
                         &  \multirow{5}{*}{0.93 , 0.93} &  0.1 &           28.7 &           28.7 &           28.7 &           17.5 &           14.9 &           36.6 &           33.8 &           12.6 &           11.6 &           46.8 &           16.2 &           10.7 &           14.9 &            9.9 &           24.3 \\
                         &                               &  0.3 &           28.7 &           28.7 &           28.7 &           16.2 &           13.7 &           36.6 &           33.8 &           11.6 &            9.9 &           46.8 &           14.9 &            7.7 &           14.9 &            7.7 &           24.3 \\
                         &                               &  0.5 &           31.1 &           31.1 &           33.8 &           36.6 &           33.8 &           39.8 &           39.8 &           19.0 &           39.8 &           50.8 &           20.7 &           19.0 &           20.7 &           19.0 &           31.1 \\
                         &                               &  0.7 &           36.6 &           36.6 &           36.6 &           46.8 &           36.6 &           43.2 &           43.2 &           50.8 &           43.2 &           50.8 &           43.2 &           50.8 &           43.2 &           50.8 &           39.8 \\
                         &                               &  0.9 &           50.8 &           50.8 &           50.8 &           50.8 &           50.8 &           50.8 &           50.8 &           50.8 &           50.8 &           50.8 &           50.8 &           50.8 &           50.8 &           50.8 &           50.8 \\
                         \cline{3-18}
                         &                          mean &      &  \textbf{31.9} &  \textbf{31.2} &  \textbf{32.1} &  \textbf{25.6} &  \textbf{21.6} &  \textbf{34.1} &  \textbf{35.4} &  \textbf{19.2} &  \textbf{18.3} &  \textbf{46.1} &  \textbf{17.9} &  \textbf{16.4} &  \textbf{18.3} &  \textbf{16.5} &  \textbf{27.9} \\
\midrule
\multirow{30}{*}{merger} &  \multirow{5}{*}{0.36 , 0.35} &  0.1 &           43.2 &           36.6 &           36.6 &            5.1 &            4.0 &           39.8 &           39.8 &            4.0 &            4.0 &           50.8 &            4.0 &            4.0 &            5.6 &            4.7 &           24.3 \\
                         &                               &  0.3 &           43.2 &           36.6 &           36.6 &           17.5 &            4.7 &           39.8 &           39.8 &            4.0 &            4.0 &           50.8 &            4.4 &            4.0 &            5.6 &            4.7 &           24.3 \\
                         &                               &  0.5 &           39.8 &           33.8 &           33.8 &           14.9 &            4.7 &           36.6 &           36.6 &            4.7 &            4.7 &           50.8 &            4.7 &            4.7 &            5.6 &            4.7 &           24.3 \\
                         &                               &  0.7 &           43.2 &           36.6 &           39.8 &           22.4 &           43.2 &           39.8 &           39.8 &           13.7 &            8.4 &           50.8 &           14.9 &            8.4 &           17.5 &           17.5 &           28.7 \\
                         &                               &  0.9 &           50.8 &           46.8 &           46.8 &           50.8 &           50.8 &           50.8 &           50.8 &           50.8 &           50.8 &           50.8 &           43.2 &           50.8 &           43.2 &           50.8 &           43.2 \\
                         \cline{3-18}
                         &  \multirow{5}{*}{0.29 , 0.27} &  0.1 &           43.2 &           36.6 &           36.6 &           17.5 &            8.4 &           39.8 &           39.8 &            5.1 &            4.7 &           50.8 &            5.1 &            4.7 &           13.7 &            5.6 &           28.7 \\
                         &                               &  0.3 &           43.2 &           36.6 &           36.6 &           19.0 &           43.2 &           39.8 &           39.8 &            6.0 &            5.6 &           50.8 &            6.0 &            5.6 &           14.9 &            7.7 &           28.7 \\
                         &                               &  0.5 &           43.2 &           36.6 &           36.6 &           19.0 &           14.9 &           39.8 &           39.8 &            6.0 &            5.6 &           50.8 &            6.0 &            5.6 &           13.7 &            8.4 &           24.3 \\
                         &                               &  0.7 &           50.8 &           43.2 &           43.2 &           28.7 &           46.8 &           46.8 &           43.2 &           24.3 &           46.8 &           50.8 &           22.4 &           50.8 &           24.3 &           50.8 &           39.8 \\
                         &                               &  0.9 &           50.8 &           50.8 &           50.8 &           50.8 &           50.8 &           50.8 &           50.8 &           50.8 &           50.8 &           50.8 &           50.8 &           50.8 &           50.8 &           50.8 &           50.8 \\
                         \cline{3-18}
                         &  \multirow{5}{*}{0.43 , 0.43} &  0.1 &           43.2 &           39.8 &           39.8 &            4.4 &            4.0 &           39.8 &           39.8 &            4.0 &            4.0 &           50.8 &            4.0 &            4.0 &            4.7 &            4.7 &           28.7 \\
                         &                               &  0.3 &           43.2 &           36.6 &           36.6 &            4.0 &            4.0 &           39.8 &           39.8 &            4.0 &            4.0 &           50.8 &            4.0 &            4.0 &            4.7 &            4.7 &           24.3 \\
                         &                               &  0.5 &           43.2 &           36.6 &           36.6 &            4.4 &            4.0 &           39.8 &           36.6 &            4.0 &            4.0 &           50.8 &            4.0 &            4.0 &            4.7 &            4.7 &           24.3 \\
                         &                               &  0.7 &           39.8 &           33.8 &           33.8 &            4.0 &            4.0 &           36.6 &           36.6 &            4.0 &            4.0 &           50.8 &            4.0 &            4.0 &            4.7 &            4.0 &           24.3 \\
                         &                               &  0.9 &           46.8 &           39.8 &           39.8 &           22.4 &           43.2 &           43.2 &           39.8 &           13.7 &            9.9 &           50.8 &           14.9 &            9.1 &           17.5 &           19.0 &           33.8 \\
                         \cline{3-18}
                         &  \multirow{5}{*}{0.70 , 0.70} &  0.1 &           22.4 &           28.7 &           28.7 &            4.0 &            4.0 &           31.1 &           33.8 &            4.0 &            4.0 &           46.8 &            4.0 &            4.0 &            4.7 &            4.0 &           22.4 \\
                         &                               &  0.3 &           22.4 &           28.7 &           28.7 &           10.7 &            5.1 &           31.1 &           33.8 &            4.7 &            4.7 &           46.8 &            5.1 &            4.7 &            5.6 &            4.7 &           22.4 \\
                         &                               &  0.5 &           20.7 &           22.4 &           24.3 &            4.4 &            4.0 &           28.7 &           31.1 &            4.0 &            4.0 &           46.8 &            4.0 &            4.0 &            4.7 &            4.7 &           20.7 \\
                         &                               &  0.7 &           31.1 &           33.8 &           33.8 &           19.0 &           20.7 &           36.6 &           36.6 &           11.6 &            8.4 &           50.8 &           11.6 &            8.4 &           14.9 &           11.6 &           24.3 \\
                         &                               &  0.9 &           50.8 &           46.8 &           46.8 &           50.8 &           46.8 &           50.8 &           46.8 &           50.8 &           50.8 &           50.8 &           43.2 &           50.8 &           43.2 &           50.8 &           43.2 \\
                         \cline{3-18}
                         &  \multirow{5}{*}{0.57 , 0.58} &  0.1 &           36.6 &           33.8 &           33.8 &            4.0 &            4.0 &           36.6 &           36.6 &            4.0 &            4.0 &           50.8 &            4.0 &            4.0 &            4.7 &            4.0 &           24.3 \\
                         &                               &  0.3 &           39.8 &           33.8 &           33.8 &            4.0 &            4.0 &           39.8 &           36.6 &            4.0 &            4.0 &           50.8 &            4.0 &            4.0 &            4.7 &            4.7 &           24.3 \\
                         &                               &  0.5 &           24.3 &           31.1 &           31.1 &            4.0 &            4.0 &           33.8 &           33.8 &            4.0 &            4.0 &           50.8 &            4.0 &            4.0 &            4.7 &            4.0 &           22.4 \\
                         &                               &  0.7 &           20.7 &           24.3 &           26.4 &            4.0 &            4.0 &           28.7 &           33.8 &            4.0 &            4.0 &           46.8 &            4.0 &            4.0 &            4.7 &            4.0 &           20.7 \\
                         &                               &  0.9 &           19.0 &           20.7 &           20.7 &            4.4 &            4.0 &           26.4 &           31.1 &            4.0 &            4.0 &           43.2 &            4.0 &            4.0 &            4.7 &            4.7 &           20.7 \\
                         \cline{3-18}
                         &  \multirow{5}{*}{0.93 , 0.93} &  0.1 &           31.1 &           33.8 &           33.8 &           17.5 &           13.7 &           36.6 &           36.6 &            7.7 &            7.1 &           50.8 &            7.7 &            6.6 &           13.7 &            8.4 &           24.3 \\
                         &                               &  0.3 &           26.4 &           31.1 &           31.1 &           16.2 &            9.9 &           33.8 &           36.6 &            6.0 &            5.6 &           50.8 &            6.0 &            5.6 &           12.6 &            7.1 &           22.4 \\
                         &                               &  0.5 &           46.8 &           39.8 &           39.8 &           33.8 &           43.2 &           43.2 &           39.8 &           19.0 &           43.2 &           50.8 &           20.7 &           22.4 &           20.7 &           26.4 &           33.8 \\
                         &                               &  0.7 &           46.8 &           43.2 &           43.2 &           50.8 &           46.8 &           46.8 &           46.8 &           50.8 &           46.8 &           50.8 &           39.8 &           50.8 &           39.8 &           50.8 &           43.2 \\
                         &                               &  0.9 &           50.8 &           50.8 &           50.8 &           50.8 &           50.8 &           50.8 &           50.8 &           50.8 &           50.8 &           50.8 &           50.8 &           50.8 &           50.8 &           50.8 &           50.8 \\
                         \cline{3-18}
                         &                          mean &      &  \textbf{38.6} &  \textbf{36.1} &  \textbf{36.4} &  \textbf{18.8} &  \textbf{19.9} &  \textbf{39.2} &  \textbf{39.2} &  \textbf{14.3} &  \textbf{15.2} &  \textbf{50.0} &  \textbf{13.5} &  \textbf{14.8} &  \textbf{15.5} &  \textbf{16.1} &  \textbf{29.1} \\
\midrule
\bottomrule
\end{tabular}
\label{tab:su and merger achromatic phase}
\caption{Duration of the achromatic phase $t_\mathrm{achro}$ in rest-frame days for all 15 LSST color curves for the  \su \, and the merger model and 30 different microlensing magnification maps ($\kappa, \gamma,$ and $s$) as in Section \ref{sec: SNe Ia models achromatic phase investigation}, where for each map 10000 random positions are drawn. The mean values are plotted in Figure \ref{fig:color curves from LSST for different supernova model}.}
\end{table}

\clearpage

\section{Additional color curves}
\label{sec:Appendix Redshifted color curves}

\begin{figure}[h]
\centering
\includegraphics[scale=0.75]{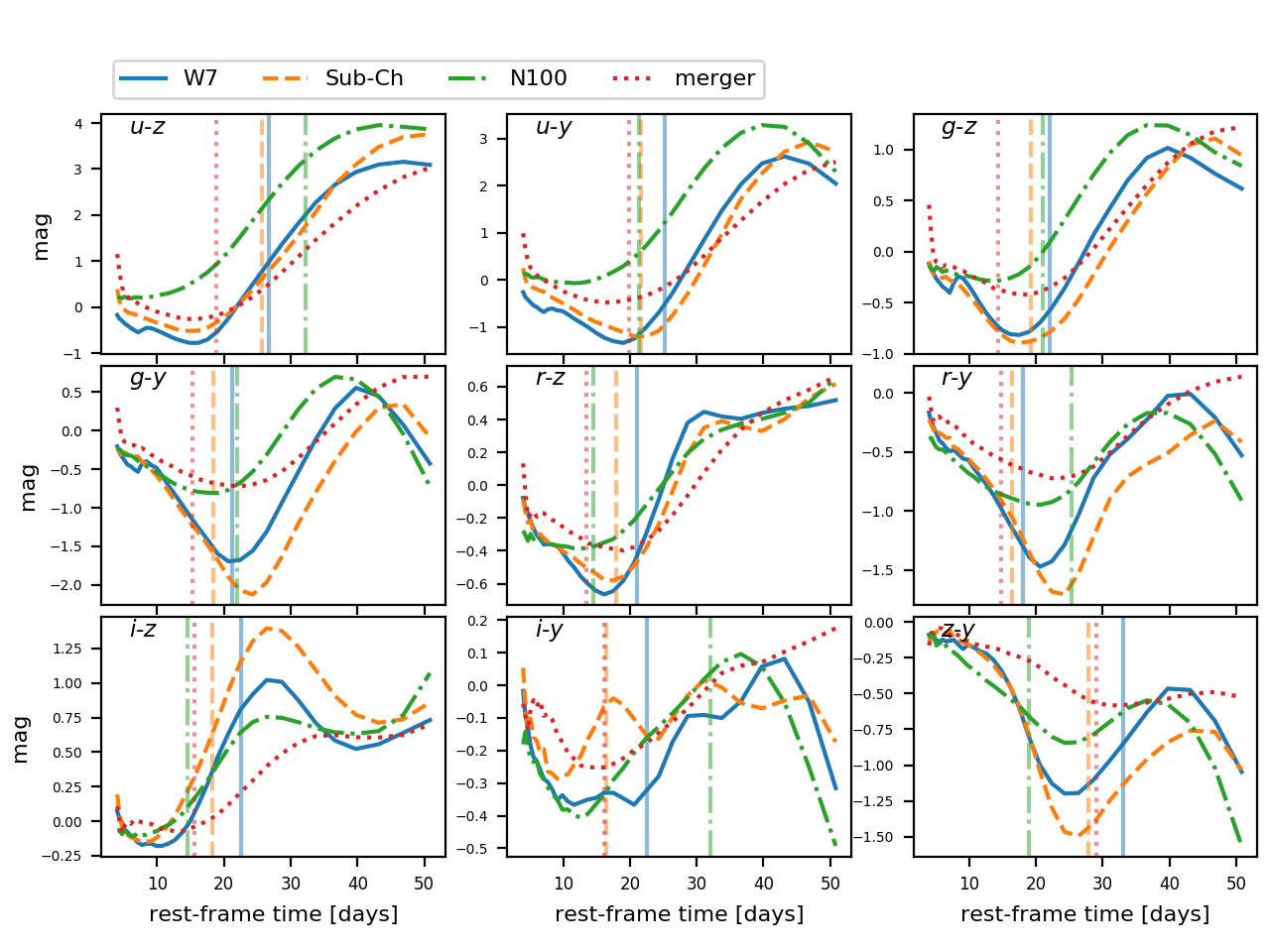}
\caption{Rest-frame LSST color curves without microlensing for
  four different SN Ia models similar as in Figure \ref{fig:color curves from LSST for different supernova model}, but this time showing the remaining nine color curves that are not as useful for time-delay measurements.}
\label{fig:color curves from LSST for different supernova model not useful}
\end{figure}

\begin{figure}[h]
\centering
\includegraphics[scale=1]{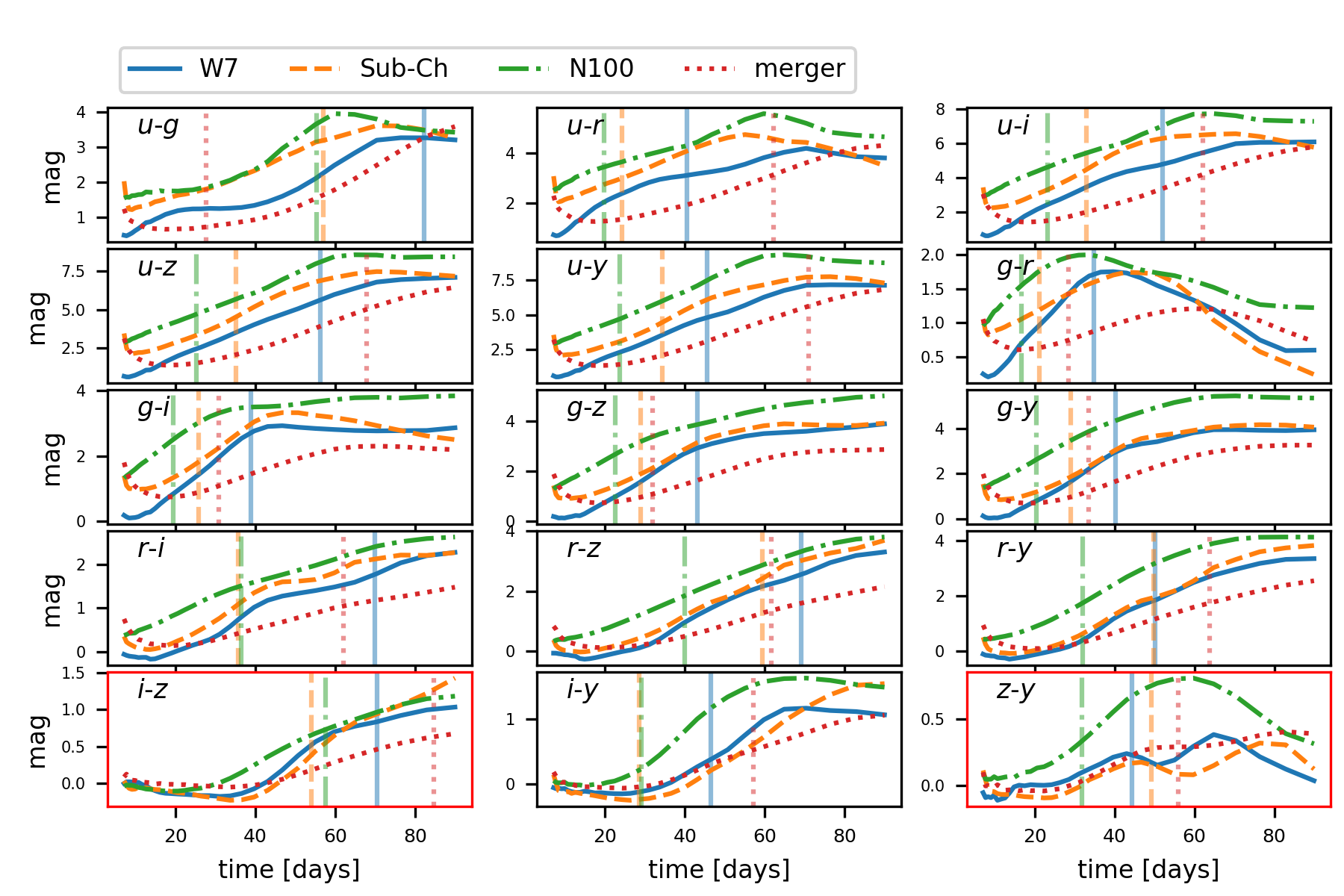}
\caption{All redshifted LSST color curves without microlensing for four
  different SN Ia models similar to Figure \ref{fig:color curves
    redshifted by 0.55} but at $\sourcez = 0.77$.}
\label{fig:color curves redshifted by 0.77}
\end{figure}

\begin{figure}[h]
\centering
\includegraphics[scale=1]{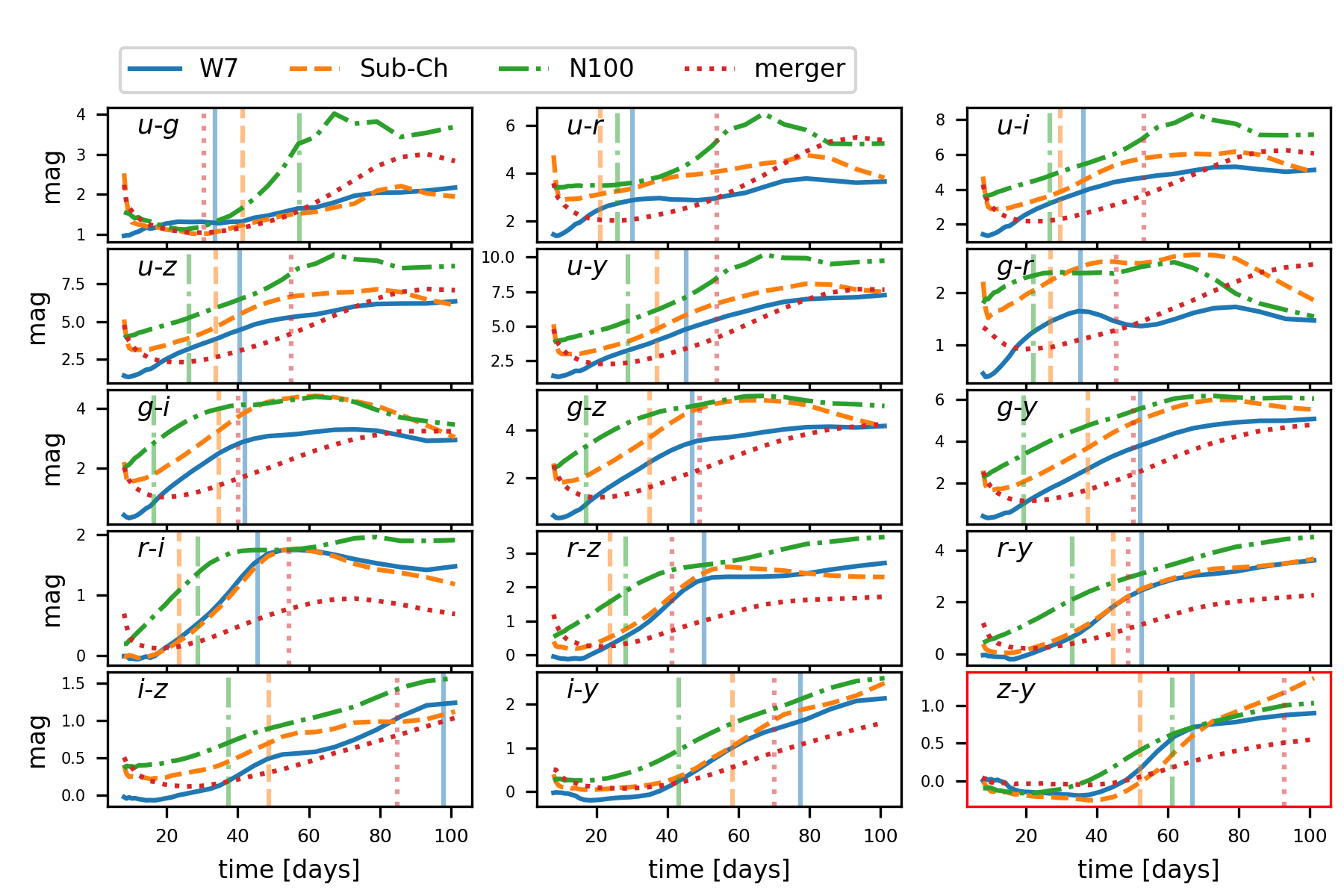}
\caption{All redshifted LSST color curves without microlensing for four
  different SN Ia models similar to Figure \ref{fig:color curves
    redshifted by 0.55} but at $\sourcez = 0.99$.}
\label{fig:color curves redshifted by 0.99}
\end{figure}

\clearpage

\end{document}